\documentclass[pdftex,twocolumn,epjc3]{svjour3}          

\RequirePackage[T1]{fontenc}

\smartqed  
\usepackage{multirow}
\RequirePackage{graphicx,subfigure}
\RequirePackage{mathptmx}      
\RequirePackage{flushend}
\RequirePackage[numbers,sort&compress]{natbib}
\RequirePackage[colorlinks,citecolor=blue,urlcolor=blue,linkcolor=blue]{hyperref}
\newcommand{\X}{\bar{X}}

\usepackage{booktabs}
\usepackage{chngpage}
\usepackage{float}

\graphicspath{{figures/}}

\journalname{Eur. Phys. J. C}

\begin{document}

\title{Geometrical aspects on the dark matter problem\thanksref{t1}}


\author{A.J.S Capistrano\thanksref{e1,addr1}
        \and
        L.A. Cabral\thanksref{e2,addr2}}

\thankstext[$\star$]{t1}{Geometrical aspects on the dark matter problem}
\thankstext{e1}{e-mail: abraao.capistrano@unila.edu.br}
\thankstext{e2}{e-mail: cabral@uft.edu.br}

\institute{Federal University of Latin-American Integration, P.O. Box 2123,85867-970, Foz do Iguaçu-PR, Brazil.\label{addr1}
          \and
          Federal University of Tocantins, P.O.Box 132, 77804-970, Araguaina-TO, Brazil\label{addr2}
          }

\date{Received: date / Accepted: date}

\maketitle

\begin{abstract}
In the present  paper  we apply  the Nash's  theory  of perturbative  geometry  to the  study  of  dark matter gravity  in a  higher-dimensional space-time. It is  shown that  the dark matter gravitational perturbations at local scale  can be  explained by   the  extrinsic  curvature  of  the  standard  cosmology. In order to test our model, we use a spherically symmetric metric embedded in a five-dimensional bulk. As a result, considering a sample of 10 low surface brightness and 6 high surface brightness galaxies, we find a very good agreement with the observed rotation curves of smooth hybrid alpha-HI measurements.
\end{abstract}

\section{Introduction}
Since the first observations of an apparently additional gra-\hspace{1mm}vitational effect for the anomalous movement of COMA cluster in 1930 by Zwicky \cite{Zwicky} and in the 70's with the rotation curve discrepancy \cite{rubin}, dark matter has been a long stand problem in cosmology and particle physics. Due to the most physical properties remain unknown, the dark matter problem has motivated several attempts to obtain a sought-after well based explanation. Interesting reviews  and a check list  for  dark matter candidates and some models can  be  found in \cite{Milgrom,Moffat,Bekenstein,Freese,Skordis,Dodelson,Mannheim,Mukhanov,Padmanabham,Hooper,Marco,capo}.

As  for  its  origin, the current  thinking is  that dark matter  may have appeared in the  very early  universe,  at the  inflationary period and that its gravitational field induced the perturbations of  matter density,  eventually producing today's  observed  large  structures  such  as   galaxies  and clusters. Hence, gravitational perturbation  theory  has  been playing an  essential  role   in our  understanding of  how the  universe   came to  be populated  by  stars and clusters  and other large  structures. The presently accepted  theory   is that  dark matter   was the cause  of that  perturbation,   but   it is not too clear  how  dark matter appeared  in the first  place.  As it is well known, perturbation mechanisms in relativistic cosmology are plagued by coordinate gauges, inherited from the group of diffeomorphism of general relativity, making it difficult to differentiate between a physical perturbation from those resulting from a mere coordinate transformation. In spite of its successes in
cosmology, the traditional gravitational perturbation theory neglects the fact that it also means perturbations
(or deformations) of the space-time geometry by a continuous sequence of local infinitesimal increments of the
metric.

In this paper we  try to avoid considerations on  exotic  dark matter  and  concentrate  only on the understanding  of  the dark matter gravitational  field and  how  it  perturbs the  gravitational  field  of  baryons. We show that at least in local scale dark matter can be neglected in the rotation curves problem, regardless of any attributes it may possess.

A concept  of smooth  perturbations is introduced based on the   generation  of  geometries  by continuous perturbation introduced by John Nash \cite{Nash} in 1956. Due to  the importance of this  concept in  cosmology,  we naturally  start  at  the  cosmological scale.  Afterwards, we see how such perturbations affect the local structures. It is also  shown that Nash's embedding theorem essentially translates the original brane-world program as a embedded space-time. Despite its very common use in most brane-world models, particular junction conditions are not considered in this paper.

This paper is organized as follows. In the  next  section,  we present the smooth perturbation based on Nash's theorem \cite{Nash,GDE}. The section 3 is devoted to discussion of the standard Friedman-Lemaître-Robertson-Walker (FLRW) cosmology, regarded as   submanifold embedded in a  five dimensional  deSitter bulk that can be  perturbed  leading  to  a  modified  Friedman's  equation. In section 4  we  apply the same  procedure to a four dimensional spherically symmetric geometry embedded in the five dimensional space-time (the bulk) and obtain the rotation curves of a sample that contains 10 low surface brightness (LSB) galaxies and 6 high surface brightness (HSB) galaxies. Final comments and foresee future works are presented in the conclusion section.

\section{Smooth perturbations}
As it  seems   evident,  the  study of dark matter gravity  and its  implications to the formation of  structure  in the early universe must  naturally  start   with gravitational perturbation theory.   The  traditional  perturbation mechanisms in relativistic  cosmology  are  plagued  by coordinate  gauges,  which are   inherited  from the group of diffeomorphism  of  general  relativity. These gauges  make it   difficult to  differentiate  between a true physical perturbation from those resulting from a mere   coordinate  transformation. Fortunately   there  are some     very  successful    criteria to  filter out the latter perturbations    in  cosmology  \cite{Bardeen,Geroch,Walker}  but they still depend on  a  choice of a  model. A lesser known, but  far   more general  approach to gravitational perturbation  was  developed   by  J. Nash,   showing   that   any  Riemannian   geometry    can be  generated by  a    continuous  sequence of local  infinitesimal   increments of a  given  geometry \cite{Nash,Greene}.

Since this  theorem  sits  at the very  foundation  of  how   geometrical  structures are  formed and  compared, it seems reasonable that  it must   be  somehow  related    to   the  formation of  structures  in  cosmology.  In principle,  these  deformations  should  be associated  with shapes  or  forms  of  large  structures  in the  early  universe.  In this  case,  the  uniqueness  and  structure of  those  forms  have  to do  with Riemann's  geometry. Rather  than being just  semantics,  this   touches   a  fundamental issue, namely the   lack  of  precision  of  Riemannian  geometry   to  determine the  curvature  of  a  manifold  endowed  with a metric  geometry.

Nash's  theorem   solves    an old  dilemma     of  Riemannian geometry,  namely that the Riemann tensor   is not sufficient  to  make  a precise  statement    about  the local shape  of  a  geometrical object or    a manifold.
The simplest   example is  given  by a 2-dimensional
Riemannian  manifold,  where  the Riemann  tensor  has only  one  component $R_{1212}$  which  coincides with  the  Gaussian  curvature. Thus,   a  flat Riemannian    2-manifold     defined by  $R_{1212}=0$   may be interpreted as
a plane,  a cylinder or a   even  a helicoid,  in the sense of  Euclidean geometry,
which is the  basis  of our  astronomical   observations.
Of course,  this ambiguity was  known   by Riemann,   who  regarded   the concept of  flatness  as defining an  equivalent class of  manifolds instead of  an  specific  one \cite{Riemann}.

However,  when we  try  to apply  this  concept  to    relativistic  cosmology, specially  referring to  structure formation, we require  a less ambiguous notion of shape of the observed  object. The solution to this  ambiguity problem  was originally proposed in  1873 by    Schlaefli \cite{Schlaefli},
conjecturing that if a  Riemannian  manifold   could be  embedded in  another one, then   a decision  on its  real  shape  could  be  made  by  comparing  the
Riemann  curvatures of the embedded   surface  with  the one  of the embedding  space. The  formal  solution of the  problem  took a long time  to  appear,
only after the  derivations of the conditions  that guarantee  the  embedding of any Riemannian  geometry into another,   the  well known Gauss-Codazzi-Ricci  equations of  geometry.
The   Gauss-Codazzi-Ricci   equations  are  non-linear and  difficult to  solve  in the general case. Some   simplifications were   obtained by  assuming that the   metric is    analytic in the  sense that it  is  a convergence of  a  positive power  series \cite{Cartan,Janet}. The  most  general  solution  appeared  only in  1956 with  Nash's theorem.

\subsection{The Nash's embedding theorem}

Nash's  theorem innovated the   embedding problem  by  introducing the  notion of  differentiable,  perturbative geometry:  using  a  continuous sequence  of  small perturbations  of a  simpler  embedded  geometry along  the  extra  dimensions, the  theorem  shows how to construct  any other  Riemannian manifold\footnote{It is worth stressing that this geometric perturbation method was apparently introduced  by  J. Campbell in a posthumous edition of a textbook  on differential geometry
\cite{Campbell} but, unfortunately, with by the use of analytic conditions\cite{Dahia}.}.  (a  brief,  updated description of
Nash's  result can be  found in  \cite{GDE,QBW}).  Thus,  Nash's    approach  to    geometry not  only  solves the ambiguity  problem  of
the  Riemannian   curvature, but also  gives  a prescription on  how to  construct geometrical  structures by deforming   simpler ones.

Reviewing some  basic  ideas of   Nash's  geometric perturbation theorem, we suppose  we have  an  arbitrarily   given   Riemannian  manifold  $\bar{V}_n$ with metric   $\bar{g}_{\mu\nu}$,  which  is  embedded  into   another  Riemannian  manifold  $V_D$,  the bulk  space.  Of  course,  $D$  must be
sufficiently large  to accommodate  the embedded  $V_n$ and  the bulk geometry
$\mathcal{G}_{AB}$   must be  given.  Assuming these  conditions,  we  generate  another  geometry  by  a     small perturbation
$\bar{g}_{\mu\nu} +  \delta g_{\mu\nu} $  where\footnote{ Greek  indices  $\mu,\; \nu...$
refer  to  the $n$-dimensional  embedded  geometries;  Small case Latin  indices $a, \; b...$ refer  to $N$ extra  dimensions;  Capital  Latin indices  $A,\; B...$
refer  to  the  bulk. Hereon, we adopt natural units where the speed of light $c$ is set as $c=1$.}
\begin{equation}
\delta g_{\mu\nu} =-2\bar{k}_{\mu\nu  a}\delta y^a,  \;\; a =N\!+\!1...D  \label{eq:geometricflow}
\end{equation}
where  $\delta  y^a$ denotes  an  infinitesimal parameter and
where  $\bar{k}_{\mu\nu a}$  denote the  extrinsic
curvature components of  $\bar{V}_n$ relative to  the extra dimension $y^a$. Using this  perturbation we obtain new  extrinsic  curvature $k_{\mu\nu  a}$, and  by repeating  the process we  obtain  a  continuous  sequence of  perturbations:
\begin{equation}
g_{\mu\nu}  =  \bar{g}_{\mu\nu}  +  \delta y \, \bar{k}_{\mu\nu}  + \delta^2 y \, \bar{g}^{\rho\sigma} \bar{k}_{\mu\rho}\bar{k}_{\nu\sigma}\cdots  \label{eq:geometricflow2}
\end{equation}
In this  way,  Nash  showed that  any  Riemannian  manifold  embedded in the bulk can be generated even if we have a flat space-time.

Nash's original  theorem   used  a  flat D-dimensional  Euclidean  space  but this  was   soon generalized  to  any  Riemannian  Manifold, including those  with   positive and  negative  signatures \cite{Greene}.  Although the  theorem  could be generalized to  include  perturbations  on  arbitrary  directions in the bulk,    it  would  make   interpretations  more difficult  so that we  retain  Nash's   choice of  independent orthogonal  perturbations. The embedding  apparently  introduces   fixed background  geometry as opposed to   a  completely intrinsic  and  self-contained  geometry  in  general relativity.  This  can  be  solved by  defining  the  geometry of  the  embedding  space  by  the  Einstein-Hilbert  variational principle,  which  has the  meaning  that   the  embedding  space  has the  smoothest possible  curvature.  This  is  compatible  with   Nash's  theorem  which  requires  a  differentiable  embedding  structure  \cite{QBW}. Another aspect of  Nash's  theorem is  that  the  extrinsic  curvature  are   the  generator of the  perturbations  of the gravitational  field  along the  extra  dimensions, as indicate the eqs.(\ref{eq:geometricflow}) and (\ref{eq:geometricflow2}).  The  symmetric  rank-2  tensor  structure of  the  extrinsic  curvature lends  the  physical  interpretation  of an    independent   spin-2  field on  the    embedded  space-time.

\subsection{Brane-world as an embedded space-time}

If one tries to adapt the mathematical embedding structure to a physical model, the original brane-world idea seems to comply with this intent. Brane-world  models have its root in a bi-dimensional model \cite{ruba} proposed an explanation of why the compact space in Kaluza-Klein theory was not observed. In other words, a confined 4-d observer does not verify a fifth dimension because it depends on the energy of observation. In that model, the physical space should be confined to a potential well embedded in a larger space (the bulk). In 1998, Arkani-Hamed, G. Dvali  and  S. Dimopolous  (ADD  for  short) proposed a  solution of the    hierarchy  problem    of the  fundamental  interactions\cite{ADD} giving the basis of the brane-world theory. It  contains essentially  three  fundamental  postulates:  (a)  the space-time  or   brane-world  is  an embedded differentiable
sub manifold of  another  space (the bulk) whose geometry is defined
by the Einstein-Hilbert action (therefore  this should not be  confused  with the  ``brane-worlds'' of string/M-theory);  (b) all gauge  interactions  are confined to the  four-dimensional brane-world  (this is  a   consequence of  the  Poincaré symmetry of the  electromagnetic  field  and in general  of the   dualities of  Yang-Mills  fields, which are   consistent  in  four-dimensional  space-time  only); and  (c)   gravitation  is defined by Einstein's  equations for the  bulk,  propagating along the  extra  dimensions at  \emph{tev} energy. It  follows  from  (b)  that   ordinary  matter must  also be  confined  because  it  interacts  with  the   confined  gauge  fields.

The  original ADD paper refers to   graviton probes to  the extra  dimensions, but  classically it means   that  the  bulk  is  locally   foliated  by a   family  brane-world  sub manifolds, whose  metric  depend  on the  extra-dimensional  coordinates in the bulk.

A different approach to compactification of extra-dimen-\hspace{1mm}sions was proposed in the Randall-Sundrum brane-world mo-\hspace{1mm}del \cite{RS,RS1} where the 4-d space-time is embedded in the five
dimensional anti deSitter space AdS5. In this model the extrinsic geometry of the universe is passive and reduced to a boundary condition depending only on the confined sources. Thus, the extrinsic curvature becomes an algebraic function of the energy-momentum tensor of matter confined to the four-dimensional embedded space-time well known as Israel-Darmois-Lanczos condition \cite{Israel}. Despite its success this condition is not unique \cite{Brandon} and cannot be apply to more than a noncompact dimension which is not the brane-world in general \cite{sepangi,maia2}.

\subsubsection{The Confinement condition}

The  four-dimensionality of the    space-time manifold  is  an  experimentally  established fact,  associated  with  the  Poincaré invariance of  Maxwell's  equations and  their  dualities, later   extended to all gauge fields. Consequently,   all matter  which  interacts   with these gauge  fields  for  consistency  must  also be  defined in  the  four-dimensional  space-times.  On the other  hand,  in  spite of  all  efforts  made so far,  gravitation has failed to  fit into  a  similar  gauge  interaction scheme,  so that the  gravitational  field  does  not necessarily have the same four-dimensional  limitations.

The   confinement  of  ordinary matter and  gauge  fields implies that
the  tangent  components of  the general source term $\alpha_* T^*_{AB}$  in the  bulk space must  coincide  with  the usual 4-d dimensional term $8\pi G  T_{\mu\nu}$ where   $T_{\mu\nu}$  is the   energy-momentum  tensor of the confined  sources. Admitting  only  the gravitational  interaction of  dark  matter,  the  above  discussion  implies  that  dark matter   cannot  be  confined  for  the  same  reasons  for  the  confinement  of  baryons.  In fact,  we  do not know  any  other property   of  dark matter  and its propagation  in the bulk,  except its  gravitational  field. It is  tempting to suppose that dark matter  also propagates  in the bulk,  so  that   the transverse  and/or  normal   components
of  $T^*_{AB}$  could    contribute to  an  observable  effect in  four  dimensions.  However, this  would   necessarily   lead  us  to  an exercise
on  dark matter  modeling.  For  example,    the  WIMP  model would give the normal  components  $T_{ab} =\rho_{DM}U_a U_b$. This  has  the important   consequence  that  the  diffeomorphism  symmetry   is also  confined  to the  brane-world. As a result, the confinement can be generally set as a condition on the embedding map such that
\begin{eqnarray}\label{eq:confinam}
&&8\pi G T_{\mu\nu}  =G_* Z^A_{,\mu}Z^B_{,\nu}T^*_{AB},\; \\
&&Z^A_{,\mu}\eta^B T^*_{AB}=0\;,\nonumber\\
&&\eta^A  \eta^B T^*_{AB}=0\;.\nonumber
\end{eqnarray}

According  to  Nash's  theorem   the  gravitational  field, regardless  of    where its  source is  located, it is   geometrically  perturbable  along the  extra  dimensions. All  that we  use  is Nash's   theorem  together $\;$ with  $\;\;$the$\;$  four-dimensionality$\;\;$ of$\;$ gauge  fields and  the  Einstein-Hilbert principle. Some similar approaches \cite{sepangi,sepangi1} have been developed with no need of particular junction conditions and/or with different junction conditions which led to a plethora of brane-world models widely studied in literature \cite{maeda,maartens,anderson,sahni,Tsujikawa,gong}.

\subsubsection{Brane-world equations in 5-D}
The natural  choice  for  the  bulk space is  Einstein-Hilbert  principle,  if  for no  other reason,  but  because that principle represents  a  statement  on   the  smoothness  of the embedding  space.  Admitting that  the perturbations are  smooth (differentiable) then    the  embedded  geometry  will  be  also differentiable.
This  smoothness of the    embedded geometry  was  a    primary  concern   of the    theorem.  The Einstein-Hilbert principle leads  to the  D-dimensional Einstein's  equations  for the  bulk
metric  $\mathcal{G}_{AB}$  in  arbitrary  coordinates
\begin{equation}
{\mathcal{R}}_{AB}-\frac{1}{2}\mathcal{R} \mathcal{G}_{AB}=\alpha_* T^*_{AB}\label{eq:EEbulk}
\end{equation}
 where $T^*_{AB}$ denotes the energy-momentum tensor   of the known  matter and gauge fields. The  constant  $\alpha_*$ determines the D-dimensional energy scale.

As in Kaluza-Klein   and   in the brane-world  theories,  the  embedding  space   $V_5$  has  a  metric  geometry  defined   by  the  higher-dimensional Einstein's  equations
\begin{equation}\label{BE0}
^5{\cal R}_{AB} -\frac{1}{2} \,{^5{\cal R}} {\cal G}_{AB}  =G_{*}  T^*_{AB} \;.
\end{equation}
where  $G_* $  is the  new  gravitational  constant  and  where $T^*_{AB}$ denotes the components  of  the energy-momentum tensor of the  known gauge fields and material sources.  From these  dynamical  equations   we   may  derive  the gravitational  field  in the embedded  space-times. Taking the tangent, vector and scalar components\footnote{The third gravitational equation was omitted here due to the fact that it vanishes in 5-D, but when the higher dimensional space-time is considered, one can obtain the equation $R-(K^2-H^2)+\mathcal{R}(D-5)=0$, sometimes called gravitational scalar equation.} of eq.(\ref{BE0})  and using the previous confinement conditions eq.(\ref{eq:confinam}) one can obtain (see    \cite{GDE}  for the
derivation of these  equations in the  high-dimensional case)
\begin{eqnarray}\label{BE1}
&&R_{\mu\nu}-\frac{1}{2}R g_{\mu\nu}-Q_{\mu\nu}=
8\pi G T_{\mu\nu}\; \hspace{2mm}\\
&&\label{BE2} k_{\mu;\rho}^{\;\rho}-h_{,\mu} =0\;,\hspace{4,9cm}
\end{eqnarray}
where  the  term   $Q_{\mu\nu}$   in the  first  equation results  from the  expression  of  $R_{AB}$ in eq.(\ref{BE0}),  involving  the  orthogonal  and  mixed    components  of  the  Christoffel  symbols for  the metric  ${\mathcal G}_{AB}$.  Explicitly  this new  term is
\begin{equation}\label{qmunu}
Q_{\mu\nu}=g^{\rho\sigma}k_{\mu\rho }k_{\nu\sigma}- k_{\mu\nu }h -\frac{1}{2}\left(K^2-h^2\right)g_{\mu\nu}\;,
\end{equation}
where     $h^2= g^{\mu\nu}k_{\mu\nu}$ is  the squared mean curvature and
 $K^{2}=k^{\mu\nu}k_{\mu\nu}$  is  the squared  Gauss curvature.
This   quantity   is  therefore   entirely  geometrical   and  it  is  conserved in the sense of
\begin{equation}\label{cons}
Q^{\mu\nu}{}_{;\nu}=0\;.
\end{equation}

It is important to stress that the  Israel-Darmois-Lanczos  condition  does not  follow  from  Einstein's  equations by  themselves. To see how it works,
consider  a  Riemannian  manifold  $\bar{V_n}$ with  metric $\bar{g}_{\mu\nu}$,  and its  local  isometric  embedding in a   D-dimensional  Riemannian  manifold   $V_D$,  $D= n+ 1$, given  by  a   differentiable  and  regular map $\bar{X}: \bar{V}_n \rightarrow  V_D$  satisfying the  embedding  equations
\begin{equation}
\mathcal{X}^{A}_{,\mu} \mathcal{X}^{B}_{,\nu}\mathcal{G}_{AB}=g_{\mu\nu},\;  \mathcal{X}^{A}_{,\mu}\bar{\eta}^B \mathcal{G}_{AB}=0,  \;  \bar{\eta}^A \bar{\eta}^B \label{eq:X}\mathcal{G}_{AB}=1,\;
\end{equation}
 where  $A,B = 1..D$ and we have  denoted by   $\mathcal{G}_{AB}$  the metric components of  $V_D$  in  arbitrary  coordinates,  and where  $\bar{\eta}$  denotes  the unit  vector  field   orthogonal  to  $\bar{V}_n$. The   extrinsic  curvature of  $\bar{V}_n$  is by  definition  the   projection of  the  variation of $\eta$   on the tangent plane \cite{eisenhart}:
\begin{equation}
\bar{k}{}_{\mu\nu} =  -\bar{\mathcal{X}}^A_{,\mu}\bar{\eta}^B_{,\nu}  \mathcal{G}_{AB}=\mathcal{X}^A_{,\mu\nu}\bar{\eta}^B \mathcal{G}_{AB} \label{eq:extrinsic}
\end{equation}
The  integration of the   system of  equations  eq.(\ref{eq:X})   gives  the   required   embedding  map  $\bar{X}$.

Construct   the    one-parameter  group of  diffeomorphism  defined    by the  map   $h_{y}(p): V_D\rightarrow V_D$,  describing  a  continuous curve    $\alpha(y)=h_y (p)$, passing  through  the  point  $p \in \bar{V_n}$,  with  unit    normal  vector  $\alpha'(p) =\eta(p)$   \cite{Crampin}.   The   group  is  characterized by   the composition  $h_{y} \circ h_{\pm y'}(p)\stackrel{def}{=} h_{y \pm y'}(p)$,  $h_{0}(p)\stackrel{def}{=}p$.
Applying   this    diffeomorphism    to   all points of  a
small  neighborhood of  $p$,  we  obtain  a   congruence   of  curves (or orbits)   orthogonal  to   $\bar{V}_n$.
It   does  not matter  if   the parameter  $y$  is  time-like or  not,   nor  if  it is  positive or  negative.

Given   a  geometric  object $\bar{\omega}$   in  $\bar{V}_n$,  its Lie  transport along the  flow   for  a small distance  $\delta y$    is  given  by     $\Omega   = \bar{\Omega}  + \delta y  \pounds_\eta{\bar{\Omega}}$,  where $\pounds_\eta$  denotes the
Lie  derivative  with respect  to   $\eta$ \cite{Crampin}.   In particular,   the  Lie  transport of the  Gaussian frame     $\{\X^A_\mu ,  \bar{\eta}^A_a  \}$, defined on  $\bar{V}_n$   gives
\begin{eqnarray}
Z^A{}_{,\mu}  &=&  X^A{}_{,\mu}  +   \delta y \;\pounds_\eta{X^A{}_{,\mu}}
=  X^A{}_{,\mu} + \delta y \;  \eta^A{}_{,\mu}\label{eq:pertu1}\\
  \eta^A  &=&\bar{\eta}^A  +   \delta y\;[\bar{\eta}, \bar{\eta}]^A
\;\;\;\;\;\;= \;\;\bar{\eta}^A \label{eq:pertu2}
\end{eqnarray}
However,  from  eq.(\ref{eq:extrinsic})  we note  that     in general   $\eta_{,\mu}  \neq \bar{\eta}_{,\mu}$.

The  set of  coordinates  $Z^A$   obtained by   integrating  these   equations  \emph{does not necessarily   describe  another  manifold}.  In order  to be  so,  they need  to  satisfy embedding  equations similar  to  eq.(\ref{eq:X}):
\begin{equation}\label{eq:embeddingeqs2}
\mathcal{Z}^A_{,\mu} \mathcal{Z}^B_{,\nu} \mathcal{G}_{AB}=g_{\mu\nu},\;  \mathcal{Z}^A_{,\mu}\eta^B \mathcal{G}_{AB}=0,  \;  \eta^A \eta^B \mathcal{G}_{AB}=1\;.
\end{equation}
Replacing  eq.(\ref{eq:pertu1}) and   eq.(\ref{eq:pertu2}) in  eq.(\ref{eq:embeddingeqs2}) and  using  the  definition  eq.(\ref{eq:extrinsic}) we obtain  the    metric  and  extrinsic  curvature of  the   new  manifold
\begin{eqnarray}
&&g_{\mu\nu} =   \bar{g}_{\mu\nu}-2y \bar{k}_{\mu\nu} + y^2 \bar{g}^{\rho\sigma}\bar{k}_{\mu\rho}\bar{k}_{\nu\sigma}\label{eq:g}\\
&&k_{\mu\nu}  =\bar{k}_{\mu\nu}  -2y \bar{g}^{\rho\sigma} \bar{k}_{\mu\rho}\bar{k}_{\nu\sigma}  \label{eq:k1}
\end{eqnarray}
Taking  the  derivative  of  eq.(\ref{eq:g}) with  respect  to  $y$ we  obtain  Nash's  deformation   condition  eq.(\ref{eq:geometricflow}).

Now we can write Einstein's   equations as
\begin{equation}\label{eq:EERS}
^5\mathcal{R}_{AB}=  G_* \left(T^*_{AB}  -\frac{1}{3}  T^* \mathcal{G}_{AB}\right)\;.
\end{equation}
The  Ricci tensor in five-dimension  $^5{\cal R}_{AB}$  may  be  evaluated  in the embedded  space-time  by  contracting it  with  the Gaussian frame $Z^A_{,\mu},\; Z^B_{,\nu}$,   $Z^A_{\mu} \eta^B$   and   $\eta^A \eta^B$. Using   eq.(\ref{eq:geometricflow}),  eq.(\ref{eq:embeddingeqs2})  and the  confinement  conditions eq.(\ref{eq:confinam}),  Einstein's  equations in eq.(\ref{eq:EERS}) become
\begin{eqnarray}
&& ^5{\cal R}_{\mu\nu}=  R_{\mu\nu}  + \frac{\partial  k_{\mu\nu}}{\partial y}  -2k_{\mu\rho}k^{\rho}_{\nu}  +  hh_{\mu\nu}  \label{eq:Ricciembedded}\\
&&  ^5{\cal R}_{\mu 5}= k_{\mu;\rho}^\rho  +
\frac{\partial \Gamma_{\mu 5}^\rho}{\partial  y}
\end{eqnarray}

In this sense, it  becomes  necessary  that  the  embedded  geometry  satisfies  particular  conditions such   that   Ricci  curvature of  the  embedding  space  coincides  with the extrinsic  curvature  of  the  embedded  space-time,   that is $^5{\cal R}_{\mu\nu}  =k_{\mu\nu}$,  which  is  not  generally  true.  One  of these  conditions is  that  the  embedded space-time  acts  as  a mirror boundary   between  two  regions  of  the  embedding  space (see  e.g. \cite{Israel}). In  this  case  we  may  evaluate the  difference  of  $^5{\cal R}_{\mu\nu}$  from  both  sides of  the  space-times  and  the  above  mentioned  boundary  condition  holds.  However,   in  doing  so  the  deformation  given  by  eq.(\ref{eq:geometricflow})  ceases  to be. Therefore,  to  find  the  deformations caused  by  the  extrinsic  curvature  such   special   conditions   are  not  applied  and   they are not needed. A more detailed discussion can be found in \cite{GDE,capis}.

\section{The cosmological model}
We naturally start at the cosmological scale. In  coordinates $(x^1,
x^2, x^3, x^4) = ( r, \theta, \phi , t )$ the FLRW  cosmological
model  can be expressed as \footnote{This is  the  same  as in  \cite{GDE}, but  written in a better known   coordinate  system, where $dr^2 \rightarrow dr^2/(1-k r^2)$  and   $ f(r)^2\rightarrow r^2$.}:
\begin{equation}
ds^2  =-dt^2  + a^2[\frac{dr^2}{1-k r^2} +
r^2(d\theta ^{2}+ \mbox{sin}^{2}\theta d\phi ^{2})]\label{eq:FLRW}
\end{equation}
In coordinates $( r, \theta, \phi , t )$ the Friedman-Robertson-Walker\\ (FLRW) model can be expressed as \cite{Rosen}:
$$ds^{2}=-dt^{2}+a^{2}(t)[dr^{2}+f_k^{2}(r)(d\theta^{2}+ \mbox{sin}^{2}\theta d\phi ^{2})]$$
where $f_k(r) = r, \mbox{sin}\, r, \mbox{sinh}\, r $
corresponding to $k =0, +1, -1$ (spatially flat, closed, open
respectively).

One can start  finding the  solution of  Codazzi's equations for the  FLRW metric in the
deSitter bulk
\[
k_{\mu[\nu;\rho]}  =0
\]
of   which   (\ref{BE2})  is  just  its  trace. The  general  solution
of  this  equation is
\begin{equation}
 k_{ij}=\frac{b}{a^{2}}g_{ij},\; \;\,
 k_{44}=-\frac{1}{\dot{a}}
\frac{d}{dt}\left(\frac{b}{a}\right)\;\; i,j = 1\ldots
3\label{eq:kij}
\end{equation}
Defining $B=\frac{\dot b}{b} $, we may express the  components of  $Q_{\mu\nu}$  as
\begin{eqnarray}
\label{eq:BB}
&&Q_{ij}= \frac{b^{2}}{a^{4}}\left( 2\frac{B}{H}-1\right)
g_{ij},\;\;\;Q_{44} = -\frac{3b^{2}}{a^{4}},\\
&&Q=-\frac{6b^{2}}{a^{4}} \frac{B}{H},  \;i,j =1..3
 \end{eqnarray}
where $H= \dot{a}/a$ is the usual Hubble  parameter. After replacing
in (\ref{BE1})  we obtain Friedman's equation   modified by  the  extrinsic  curvature \cite{GDE}:
\begin{equation}\label{eq:equacao de Friedman modificada}
\left(\frac{\dot{a}}{a}\right)^2+\frac{\kappa}{a^2}=\frac{8}{3}\pi
G\rho+\frac{\Lambda}{3}+\frac{b^2}{\epsilon a^4}\;\;.
\end{equation}
where $a(t)$ is the usual expansion parameter of the Universe. The same results were independently obtained by the authors in \cite{sepangi} also without assumption of any junction condition. It is important to note that the perturbed term $\frac{b^2}{\epsilon a^4}$ in eq.(\ref{eq:equacao de Friedman modificada}) is differently from that one provided by the Randall-Sundrum model with the quadratic density term that leads to a serious cosmological constraints \cite{binetruy}. In ref.(\cite{GDE}) it was shown that once the Israel-Darmois-Lanczos condition is imposed on eq.(\ref{eq:equacao de Friedman modificada}), the quadratic density term is obtained.

It was shown in \cite{gde2} that the contribution of the extrinsic curvature, $b(t)$, can be given by
\begin{equation}\label{eq:solucao geral b(t)}
b(t)= \alpha_0\left(a\right)^{\beta_0}e^{\mp \frac{1}{2}\gamma(t)}\;\;,
\end{equation}
where all integration constants were combined in $\alpha_0$ and $\beta_0$ and the term $\gamma(t)$ denoted by
\begin{equation}\label{eq:gamma}
\gamma(t)= \sqrt{4\eta_0^2a^4 - 3}-
\sqrt{3}\arctan\left(\frac{\sqrt{3}}{3}\sqrt{4\eta_0^2a^4
- 3}\right)\;.
\end{equation}

Alternatively, the modified Friedman equation can be written as
\begin{equation}\label{eq:friedman modific por gupta}
\left(\frac{\dot{a}}{a}\right)^2+\frac{\kappa}{a^2}=\frac{4}{3}\pi
G\rho+\frac{\Lambda}{3}+\kappa_0a^{2\beta_0-4}e^{\gamma(t)}
\end{equation}
where $\eta_0$ and $\kappa_0=\frac{b^2_0}{a^{2\alpha_0}_0}$ are integration constants. In addition, this  result in first approximation can be compared with  the phenomenological x-fluid model (XCDM),  with  state equation $ p_{x} = \omega_{x} \rho_{x}$,   which  corresponds  to the  geometric  equation  on $b(t)$ which in the   particular  case when   $\omega_{x}=\omega_{0}$=constant, one obtain  a simple solution
\begin{equation}
 b(t) = b_{0}(\frac{a}{a_{0}})^{\frac{1}{2}(1-3\omega_{0})} \label{eq:b}
\end{equation}
where $a_{0}$ and $b_{0} \neq 0$ are integration constants. This solution is  consistent  with  the most recent observations within the range   $-1\leq \omega_{0} \leq -1/3$ \cite{GDE, gde2}, for $\omega = -1$ results in $\Lambda$CDM cosmological model. Furthermore, the theoretical  power  spectrum   obtained  from the  extrinsic  curvature perturbation of the  FLWR model,   is  not  very different  from  the  observed  power  spectrum  in the  Planck experiment \cite{planck}.

\section{Local Dark Matter Gravity and rotation curves}
It is possible that   the  gravitational  field  of  young  galaxies  which are  still in the process of  formation \cite{Puzia};  in  galaxies    with   active galaxy  nuclei \cite{deVries};   or  even  in  cluster  collisions, Nash's perturbations  similar  to the  cosmological case  could  be  applied,  where the metric  symmetry  is  taken  to be local,  instead of the homogeneous and  isotropic  conditions  which  lead  to  the $G_2$ metric  symmetry  of the  standard  cosmological model.

These  equations  are  be understood  in the context of  the  embedded  space-times and  with the  confinement  conditions for  ordinary matter  and gauge fields. They  do not represent  the  whole of  general relativity  because the principle of   general  covariance   does not necessarily apply  to the bulk geometry. This  follows  from  the fact  that   Nash's  perturbations  are  restricted to be along  the orthogonal  directions   only.

In order  to  see  how  realistic  such  description is, we apply it and compare to the observational data. Since  in  eq.(\ref{BE1}) and  eq.(\ref{BE2}),  the metric  and  extrinsic  curvature  are  independent   variables, thus non-trivial cosmological  perturbations are  obtained  only  from a non-vanishing extrinsic  curvature.

\subsection{The model}
Consider a four dimensional spherically symmetric metric in a form
$$ds^2 = B(r) dt^2 -  A(r) dr^2 - r^2 d\theta^2 - r^2 \sin^2\theta d\phi^2\;\;.$$
We can straightforwardly obtain the components of Ricci tensor as
$$R_{rr} = \frac{B''}{2B} - \frac{1}{4}\frac{B'}{B} \left(\frac{A'}{A} + \frac{B'}{B} \right) - \frac{1}{r} \frac{A'}{A}\;,$$
$$R_{\theta\theta} = -1 + \frac{r}{2A} \left(-\frac{A'}{A} + \frac{B'}{B} \right) + \frac{1}{A}\;,$$
$$R_{\phi\phi} = \sin^2\theta R_{\theta\theta}\;,$$
and also
$$R_{tt} = -\frac{B''}{2A} + \frac{1}{4}\frac{B'}{A} \left(\frac{A'}{A} + \frac{B'}{B} \right) - \frac{1}{r} \frac{B'}{A}\;,$$
where we denote $\frac{dA}{dr} = A'$ and $\frac{dB}{dr} = B'$.

The gravi-tensor equation in five dimensions can be written as
\begin{equation}\label{eq:1}
    R_{\mu\nu} - \frac{1}{2} Q g_{\mu\nu} = Q_{\mu\nu}
\end{equation}
where $Q= g^{\mu\nu} Q_{\mu\nu}$.

The general solution of Codazzi equation can take the form
\begin{equation}
    k_{\mu\nu}=f_{\mu}g_{\mu\nu} \;\;\;\;(no\;sum\;on\;\mu)\;,\label{eqn:geneqk}
\end{equation}
where $f_{\mu}$ represents a set of scalar functions. Thus, considering the definition of $Q_{\mu\nu}$, we have:
$$Q_{\mu\nu}= f^2_{\mu}g_{\mu\nu}-\left(
\sum_{\alpha}f_{\alpha}\right)f_{\mu}g_{\mu\nu}-\frac{1}{2}\left(\sum_{\alpha}f^2_{\alpha}
-\left(\sum_{\alpha}f_{\alpha}\right)^2\right)g_{\mu\nu}\;,$$
where we identify
$$U_{\mu}=f^2_{\mu}-\left( \sum_{\alpha}f_{\alpha}\right)f_{\mu}
-\frac{1}{2}\left(\sum_{\alpha}f^2_{\alpha}-\left(\sum_{\alpha}f_{\alpha}\right)^2\right)\delta^\mu_\mu\;.$$
Thus, one can rewrite $Q_{\mu\nu}$ in terms of $f_{\mu}$ as
\begin{equation}
Q_{\mu\nu}= U_{\mu}g_{\mu\nu} \;\;\;\;\;(no\;\;sum\;\;on\;\;\mu).
\end{equation}
Since $Q_{\mu\nu}$ is a conserved quantity, once can find 4 equations from $\sum_{\nu} g^{\mu\nu}U_{\mu;\nu}=0$ which can be reduced to two equations:
$$
\left\{
  \begin{array}{ll}
    \left(f_1 + f_2\right) \left(f_3 - f_4\right) = 0 & \hbox{;} \\
    \left(f_3 + f_4\right) \left(f_1 - f_2\right) = 0 & \hbox{.}
  \end{array}
\right.
$$
and results in
$$f_1 f_3 = f_2 f_4\;.$$
We have two options: $f_1 = f_2$ e $f_3 = f_4$ $f_1 = f_3$ e $f_2 = f_4$. Taking the first option, $ f_1 = f_2 = \alpha(r)$ that leads us to $ f_3 = f_4 = \beta(r)$, where $\alpha(r)$ and $\beta(r)$ are arbitrary functions. In addition, we can find the $Q_{\mu\nu}$ components:
$$
\left\{
  \begin{array}{ll}
    K^2 = 2 \left(\alpha^2(r) + \beta^2(r)\right) & \hbox{,} \\
    h = 2 \left(\alpha(r) + \beta(r)\right) \rightarrow H^2 = 4 \left(\alpha(r) + \beta(r)\right)^2 & \hbox{,}\\
    K^2 - H^2 = -2 \left(\alpha^2(r) + \beta^2(r)\right) - 8 \alpha(r)\beta(r) & \hbox{.}
    \end{array}
\right.
$$
And we obtain  $Q_{\mu\nu} = Q_{ii} + Q_{jj}$:
\begin{equation}\label{eq:Qmunu}
\left\{
  \begin{array}{ll}
    Q_{ii} = \left(\alpha^2(r)  + 2\alpha(r)\beta(r)\right) g_{ii} & \hbox{,} \\
    \;\;\;\;\;\;\;\;\;\;\;\;\;\;\;\;\;\;\;(for\;\;(11)\;\;and\;\;(22)\;components) & \hbox{,}\\ \nonumber
    Q_{jj} = \left(\beta^2(r)  + 2\alpha(r)\beta(r)\right) g_{jj} & \hbox{,} \\
    \;\;\;\;\;\;\;\;\;\;\;\;\;\;\;\;\;\;\;(for\;\;(33)\;\;and\;\;(44)\;components).\nonumber
\end{array}
\right.
\end{equation}
and the trace
\begin{equation}\label{eq:tracoQ}
Q = 2\alpha^2(r) + 2\beta^2(r) + 8\alpha(r)\beta(r)\;.
\end{equation}

From eq.(\ref{eq:1}), we find the Ricci tensor components:
\begin{equation}\label{eq:ricci}
\left\{
  \begin{array}{ll}
    R_{ii} = \left(\alpha^2(r) + 2\beta^2(r) + 6 \alpha(r)\beta(r) \right) g_{ii} & \hbox{,} \\
    \;\;\;\;\;\;\;\;\;\;\;\;\;\;\;\;\;\;\;(for\;\;(11)\;\;and\;\;(22)\;components) & \hbox{,}\\ \nonumber
    R_{jj} = \left(\beta^2(r) + 2\alpha^2(r) + 6 \alpha(r)\beta(r) \right) g_{jj} & \hbox{,} \\
    \;\;\;\;\;\;\;\;\;\;\;\;\;\;\;\;\;\;\;(for\;\;(11)\;\;and\;\;(22)\;components) & \hbox{.}\\ \nonumber
\end{array}
\right.
\end{equation}

Using the $R_{11}$ and $R_{44}$ components, we find:
\begin{equation}\label{eq: ab}
\frac{A'}{A} + \frac{B'}{B} = \left(\beta^2 - \alpha^2\right)\;,
\end{equation}
which can be integrated as
\begin{equation}\label{eq: ab2}
AB = \exp\left( - \left[\int \left(\beta^2 - \alpha^2\right)  r dr  - C \right]\right)\;,
\end{equation}
where $C$ is a integration constant.

Using the contour
$$\lim_{r \rightarrow \infty }A(r) = \lim_{r \rightarrow \infty}  B(r)= 1\;,$$
and without loss of generality, we can set $C=0$ and find
\begin{equation}\label{eq: ab3}
A(r) = \frac{\sigma(r)}{B(r)}
\end{equation}
where we denote $\sigma(r)=\int \left(\beta^2 - \alpha^2\right)  r dr  $.

If we set the another possible option, i.e, $f_1 = f_3$ and $f_2 = f_4$, we will end up in the same situation. Thus, setting $\alpha(r) = \beta(r)$, it allows us to do
\begin{equation}\label{eq: 11}
\frac{B''}{2B} - \frac{1}{4}\frac{B'}{B} \left(\frac{A'}{A} + \frac{B'}{B} \right) - \frac{1}{r} \frac{A'}{A}= 9 \alpha^2(r)A\;.
\end{equation}
and also
\begin{equation}\label{eq:12}
-\frac{B''}{2A} + \frac{1}{4}\frac{B'}{A} \left(\frac{A'}{A} + \frac{B'}{B} \right) - \frac{1}{r} \frac{B'}{A}= -9\alpha^2(r)B\;,
\end{equation}
and one can find
$$\frac{A'}{A} =- \frac{B'}{B}\;,$$
thus,
$AB =\;constant\;.$
Taking the Minkowskian contour
$$\lim_{r \rightarrow \infty }A(r) = \lim_{r \rightarrow \infty}  B(r)= 1\;,$$
we have
$$A(r) = \frac{1}{B(r)}\;.$$

On the other hand, using the former condition, we can write the component $(\theta,\theta)$ of the Ricci tensor as
$$R_{\theta\theta}= -1+ B'r + B\;\,$$
and using(\ref{eq:12}), we find
\begin{equation}\label{eq:bint}
B(r)= 1 + \frac{K}{r} + \frac{9}{r} \int \alpha^2(r) r^2 dr\;,
\end{equation}
where $K$ is an integration constant, and also
\begin{equation}\label{eq:aint}
A(r)= \left[B(r)\right]^{-1} = \left[ 1 + \frac{K}{r} + \frac{9}{r} \int \alpha^2(r) r^2 dr\; \right]^{-1}\;.
\end{equation}

We point out that the $B(r)$ function remains undetermined due to the fact that the $\alpha(r)$ and $\beta(r)$ functions are arbitrary which do not allow the integration of $B(r)$. It is a consequence of the homogeneity of Codazzi equation (\ref{BE2}) which is not possible to be solved in 5-dimensions without an additional equation or a particular condition. Moreover, it was shown that the embedding of Schwarzschild space-time can only be possible in at least six dimensions \cite{kasner,kasner2,fronsdal,kruskal,Szekeres}. As a result, one gets a very constrained embedded geometry which can lead a serious physical consequences depending on how the extrinsic curvature is taken into account.

In order to attenuate the embedding constraint, we can try to obtain a solution with a minimum assumption with the use of the asymptotically conformal flat condition
\begin{equation}\label{eq:flatcondition}
\lim_{r\rightarrow \infty} k_{\mu\nu} = \lim_{r\rightarrow \infty} \alpha(r)\lim_{r\rightarrow \infty} g_{\mu\nu}\;.
\end{equation}
In first approximation, this condition is simply derived from the analysis on the behavior of the extrinsic curvature at infinity.\\

As $\lim_{r\rightarrow \infty} g_{\mu\nu} \rightarrow \eta_{\mu\nu}$, where $\eta_{\mu\nu}$ is $M_{4}$ metric, the extrinsic curvature vanishes as it tends to infinity, so the function $\alpha(r)$ must comply with this condition. Thus, we can infer conveniently that the function $\alpha(r)$ must be analytical at infinity such as
\begin{equation}\label{eq:alfaform}
\alpha(r)= \frac{\sqrt{|\alpha_0|}}{r^n}\;,
\end{equation}
where $\alpha_0$ is a constant (that represents the influence of extrinsic curvature) and the index $n$ represents all the set of scalar fields that fall off with $r$-coordinate following the inverse $n^{th}$ power law and $n\geq 0$ in order to comply with eq.(\ref{eq:flatcondition}).\\

Using the equations (\ref{eq:bint}) and (\ref{eq:alfaform}), one can obtain
\begin{equation}\label{eq:br2}
B(r) = 1 + \frac{K(9\alpha_0 + 1)}{r} - \frac{9\alpha_0}{3-2n} r^{2\left(1-n\right)}\;.
\end{equation}
In addition, from the eq.(\ref{eq:br2}) one can calculate the gravitational potential and find
\begin{equation}\label{eq:potencial}
\Phi(r) = -1- \frac{K \left( 9\alpha_0 + 1 \right)}{2r}+(9/2)\alpha_0 \frac{r^{(2-2n)}}{(3-2n)}\;.
\end{equation}
It is worth stressing to the reader on how the obtained gravitational potential on eq.(\ref{eq:potencial}), which without the second term it resembles the Schwarzschild solution, is affected by extrinsic curvature resulting in a modification so far of the very symmetric aspect of the original spherical symmetry. This characteristic will be decisive to deal with the rotation curve problem.

Thus, using the newtonian expression for a orbiting particle $v(r) = \sqrt{r \frac{\partial \Phi}{\partial r}}$ the velocity rotation modified by extrinsic curvature is given by
\begin{equation}\label{eq:velocrot}
v(r)=\sqrt{r | (1/2) \frac{K\left(9 \alpha_0+1 \right)}{r^2}+ (9/2) \alpha_0 \frac{(2-2n) r^{(2-2n)}}{(3-2 n) r}|}\;.
\end{equation}

In order to make a suitable notation to eq.(\ref{eq:velocrot}) we set new parameters $\beta_0$ and $\gamma_0$ with a following correspondence to parameter $n$ and $\alpha$ as
\begin{equation}\label{eq:newparameters}
\beta_0 = 2-2n\;\;\;;\;\;\;\gamma_0 = \frac{9}{2} \frac{\alpha_0 \beta_0}{(1+\beta_0)}\;.
\end{equation}

The appropriate relation to the standard gravity is set when $\frac{K \left( 9\alpha_0 + 1 \right)}{2}= GM$. Thus, the correct units for the rotation velocity in eq.(\ref{eq:velocrot}) are guaranteed.  In addition, the bulge is modeled  using Blumenthal's mass model defined by\\\vspace{0.1cm}
\begin{equation}\label{eq:mass model}
M(r)=M_{\star} \left[1 - (1+ \frac{r}{r_l}) \;\exp{\left(-\frac{r}{r_l}\right)}\right]\;.
\end{equation}
where $r_l$ is the scale length parameter and $M_{\star} = M_{gas} + M_{disk}$.
\vspace{1cm}
\begin{center}
\begin{table*}[hpt]
{\small
\hfill{}
\begin{tabular}{*{10}{l}}
\toprule
\bfseries \bfseries Galaxies & \bfseries D (Mpc)& \bfseries $R_c\;(kpc)$ & \bfseries $R_{max}\;(kpc)$ & \bfseries $M_{HI}(10^{10}M_{\bigodot})$ & \bfseries $M_{disk}(10^{10}M_{\bigodot})$ & \bfseries $V_{gas}(km/s)$ & \bfseries $V_{disk}(km/s)$ & \bfseries $V(km/s)$  & rotation curve\\
\midrule
F563-1                 &     45    & 2.9   & 17.5         & 0.29   &  1.35   & 32.8    & 20.8        & 110.9       &  fig.01 (a)      \\
F571-8                 &    48     & 5.4   & 14.0         & 0.16   & 4.48    & 33.3     & 34.6        & 143.9      &  fig.01 (b)      \\
F579-V1                &    85     & 5.2    &  14.4       & 0.21   &3.33      &  33.0      & 34.7      & 114.2     &  fig.01 (c)         \\
F583-1                 &    32     & 1.6      & 14.0       &0.18   & 0.15      & 38.2      & 9.82      & 86.9      &  fig.01 (d)        \\
UGC11454                 &    91     & 3.4    & 11.9       &  -    &  3.15     &   -    &  -          &  152.2     &  fig.02 (a)         \\
UGC11748                 &    73     & 2.6    & 21.0       &  -    &  9.67     &   -    &  -          &  246.5     &  fig.02 (b)           \\
UGC11819                 &    60     & 4.7    & 11.7       &  -    &  4.83     &   -    &  -          &  262.8     &  fig.02 (c)           \\
ESO01400040              &    212     & 10.1   & 29.9       &  -    &  20.70     &  -    &  -          &  262.8     &  fig.02 (d)         \\
ESO2060140               &  60       & 5.1    & 11.65      &  -    &  3.51     &   -    &  -          &  118.0     &  fig.02 (e)            \\
ESO4250180               &  86       & 7.3    & 14.4       &  -    &  4.79     &   -    &  -          &  144.5     &  fig.02 (f)            \\
\toprule
\end{tabular}}
\hfill{}
\caption{Relevant observed properties of 10-LSB galaxies. A larger sample of the data can be obtained in \cite{blok,Mannheim2} and also with error data in Mcgaugh's data base (http://astroweb.case.edu/ssm/data/RCsmooth.0701.dat.)}
\label{tb:tablename}
\end{table*}
\end{center}

\begin{center}
\begin{table*}[htp]
{\small
\hfill{}
\begin{tabular}{*{8}{l}}
\toprule
\bfseries \bfseries Galaxies &  $r_l$ & $\gamma_0$ & $\beta_0$ & $\chi^2_{red}$ & $p$\\
\midrule
F563-1                 &    2.844 +/- 0.471   &  0.509  +/- 0.213  &  0.263   +/- 0.238   &    0.111   &  0.9976   \\
F571-8                 &    5.702 +/- 2.703   &  1.855  +/- 1.94   &  -0.403  +/- 0.632   &    0.423   &  0.9363   \\
F579-V1                &    2.507 +/- 0.116   &  0.282  +/- 0.078  &  0.648   +/- 0.333   &    0.212   &  0.9968   \\
F583-1                 &    3.704 +/- 3.458   &  2.514  +/- 5.783  &  -0.224  +/- 0.887   &    0.160   &  0.9998  \\
UGC11454               &    4.828 +/- 5.572   &  2.609  +/- 6.508  &  -0.493  +/- 1.149   &    1.434   &  0.0455   \\
UGC11748               &    1.771 +/- 0.187   &  0.423  +/- 0.117  &  0.082   +/- 0.162   &    2.618   &  0.0005   \\
UGC11819               &    5.791 +/- 14.91   &  2.406  +/- 13.14  &   -0.345 +/- 2.820   &    0.718   &  0.7353   \\
ESO01400040            &    5.117 +/- 0.567   &  1.324  +/- 0.272  &  0.025   +/- 0.218   &    0.107   &  0.9907   \\
ESO2060140             &    3.974  +/- 7.256  &  1.238  +/- 4.873  &  -0.596  +/- 2.903   &    0.200   &  0.9984   \\
ESO4250180             &    5.863  +/- 0.712  &  1.296  +/- 0.358  &  0.263   +/- 0.263   &    0.018   &  0.9998   \\
\toprule
\end{tabular}}
\hfill{}
 \caption{Fitting parameters for 10-LSB galaxies. The $\chi^2_{red}$ and the chance $p$ are shown for each galaxy.}
\label{tb:tablename3}
\end{table*}
\end{center}

We consider here only the mass of gas and the mass of the galactic disk. The mass scale for the primordial He is given by $M_{gas} = 1.4 M_{HI}$, where $M_{HI}$ is the mass of the hydrogenous 21cm lines that go through the galactic disk to the outermost regions. As the rotation curve goes up to the bound of a circular disk one can obtain
\begin{equation}\label{eq:velocrot2}
v(r)=\sqrt{\frac{GM(r)}{r}+ \frac{GM(r)}{r_c}\gamma_0\left(\frac{r}{r_c}\right)^{\beta_0}}\;.
\end{equation}
where $r_c$ is the disk scale length.

In terms of the parameters $\beta_0$ and $\gamma_0$, it is important to consider that when the perturbation induced by the extrinsic curvature $k_{\mu\nu}$ vanishes, it implies that $\alpha_0 = 0$ and also $\gamma_0 = 0$, as indicates eq.(\ref{eq:newparameters}). In this case, we recover the Schwarzschild solution. On the other hand, when the parameter $\beta_0$ vanishes the extrinsic curvature decays as $\sim 1/r$ which is not compatible with the observed rotation curves. Moreover, to comply with eq.(\ref{eq:newparameters}),  the parameter $\beta_0$ is constrained by $\beta_0 \neq 1$ .

The case when $\beta_0=2$ is also neglected even if it induces umbilicus points as expected for a spherical geometry, but the presence of such umbilicus is inconsistent with a rotation curve solution in asymptotic regions where the dark matter dilemma arises. It resembles a Schwarzschild-deSitter solution modified by the extrinsic term with a gravitational potential in a form
$$
\Phi  =  - \frac{M}{r} + C^{\star}\,r^2\;,
$$
where $C^{\star}$ is an integration constant. Consequently, any point in the braneworld with the condition $Q=-(K^2-h^2)=0 $ or $ Q=\;constant$ must be umbilicus, which means that it has a non-trivial solution when $K=\pm h$ or $y= \mp 1$. This is not the case of a braneworld in general, and it may occur in very special situations where all directions $dx^{\mu}$ are principal directions.
\newpage
With these caveats in mind, we are able to deal with the rotation curves problem \emph{per se}.

\begin{center}
\begin{table*}[htp]
{\small
\hfill{}
\begin{tabular}{*{9}{l}}
\toprule
\bfseries \bfseries Galaxies & \bfseries D (Mpc)& \bfseries $R_c\;(kpc)$ & \bfseries $R_{max}\;(kpc)$ & \bfseries $M_{HI}(10^{10}M_{\bigodot})$ & \bfseries $M_{disk}(10^{10}M_{\bigodot})$ &   \bfseries $v(km/s)$  &  rotation curve\\
\midrule
NGC3726                 &  17.5       & 3.2   & 31.47       &0.60    & 3.82       & 167     &   fig.03 (a)         \\
NGC3769                 &  15.5       & 1.5   & 32.01       & 0.41   &  1.36      & 113     &   fig.03 (b)         \\
NGC3992                 &   25.6      & 5.7   & 49.64       & 1.94   & 13.94      & 237     &   fig.03 (c)         \\
NGC4100                 &   21.4      & 2.9   & 27.07       &0.44    &5.74        & 159     &   fig.03 (d)         \\
NGC4157                 &  18.7       & 2.6   & 30.82       &0.88    & 5.83       & 185     &   fig.03 (e)          \\
NGC4217                 & 19.6        & 3.1   & 18.15       &0.30    & 5.53       & 178     &   fig.03 (f)          \\
\toprule
\end{tabular}}
\hfill{}
 \caption{Relevant observed properties of HSB galaxies. A larger sample can be obtained in \cite{verheijen}.}
\label{tb:tablename2}
\end{table*}
\end{center}

\subsection{The samples}
The main criteria used to choose an appropriated sample was to study galaxies with at least 11kpc and also being asymptotic enough to check our model in the region where the discrepancy is most critical. To this end, we chose a sample of 10-LSB galaxies varying from 11.6kpc up to 61.9kpc. As well known, the LSB galaxies are in principle dark dominated and they provide a good test for a gravitational theory \cite{blok,Mannheim2}. The present sample was extract from \cite{blok}. The authors in \cite{blok} presented a larger sample of 30-galaxies based on the analysis of high-resolution of smooth hybrid alpha-HI rotation curves. The measurements were obtained with long-slit major axis spectra taken with the 4 m telescope at Kitt Peak in and the 2.5 m telescope at Las Campanas Observatory. The adopted distances were computed assuming Hubble constant $H_0= 75\;km^{-1}\;Mpc^{-1}$.

The table (1) shows the main values considered to fit rotation curves. The data of disk scale length $R_c$, HI gas mass $M_{HI}$ that extends beyond optical disk and the disk mass $M_{disk}$ were obtained from \cite{Mannheim2}. In addition, the adopted distances $D$, the maximum radius $R_{max}$ in kpc for adopted distance and the main velocities $V_{gas}$ (rotation caused by the observed gas), $V_{disk}$ (rotation caused by the observed disk stars) and $V$(the observed smoothed velocity) were obtained from \cite{blok}.

The rotation curves of 10-LSB galaxies are presented in fig.(1) and (2). It is worth noting that in the fig.(1) we present the theoretical and the observed rotation curves with solid lines and error bars respectively. In addition, the observed quantities of the velocity of the gas (dashed lines) extended far beyond the visible disk and the velocity caused by disk stars (starred lines) are also shown. In fig.(2) we show  both theoretical (solid lines) and observed (error bars) rotation curves. The dashed lines indicate rotation curve without extrinsic curvature.

The quantities $M_{HI}$,$V_{disk}$ and $V_{gas}$ in table (1) with respect to the following galaxies: UGC11454, $\;\;\;\;$UGC11748, $\;\;\;\;$UGC11819, ESO01400040, ESO2060140 and ESO4250180 were not observed or presented a very small amount to take into account.

In the table (2), the values of $r_l, \gamma_0, \beta_0$ are presented as well as $\chi^2_{red}$ that stands for the reduced $\chi^2$. The \emph{chance} $p$ represents the probability of concordance between the model and the data.

For the fits, we used GnuPlot 4.6 to compute non-linear least-squares fit applying the Levenberg-Maquardt algorithm. Since $r_l, \gamma_0, \beta_0$ are not universal constants, they must vary from one galaxy to another.

For the HSB galaxies we consider a different sample extracted from \cite{verheijen}. Following the same criteria as to the LSB galaxies, we study a minor sample of 6-HSB galaxies with its radii varying from 18 kpc up to 49.6 kpc in the Ursa Major cluster.

In the table (3) we present the main values considered for evaluating the rotation curves of HSB galaxies. The data of individual adopted distance $D$, disk scale length $R_c$, HI gas mass $M_{HI}$ and disk mass $M_{disk}$ were obtained from \cite{Mannheim2}.

\begin{figure*}[htb!]
     \begin{center}
        \subfigure[ ]{%

            \includegraphics[width=0.5\textwidth]{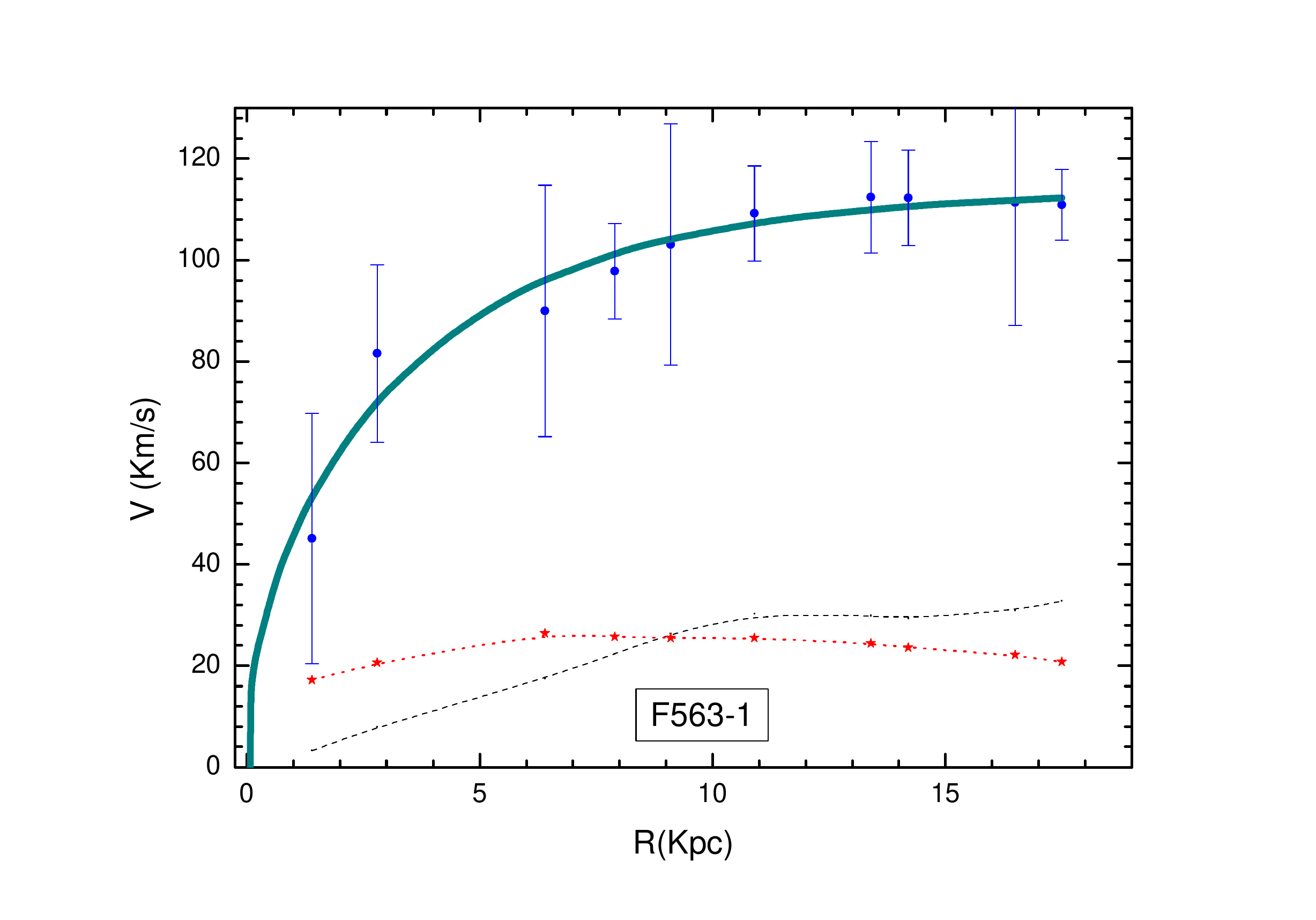}
        }%
        \subfigure[ ]{%

           \includegraphics[width=0.46\textwidth]{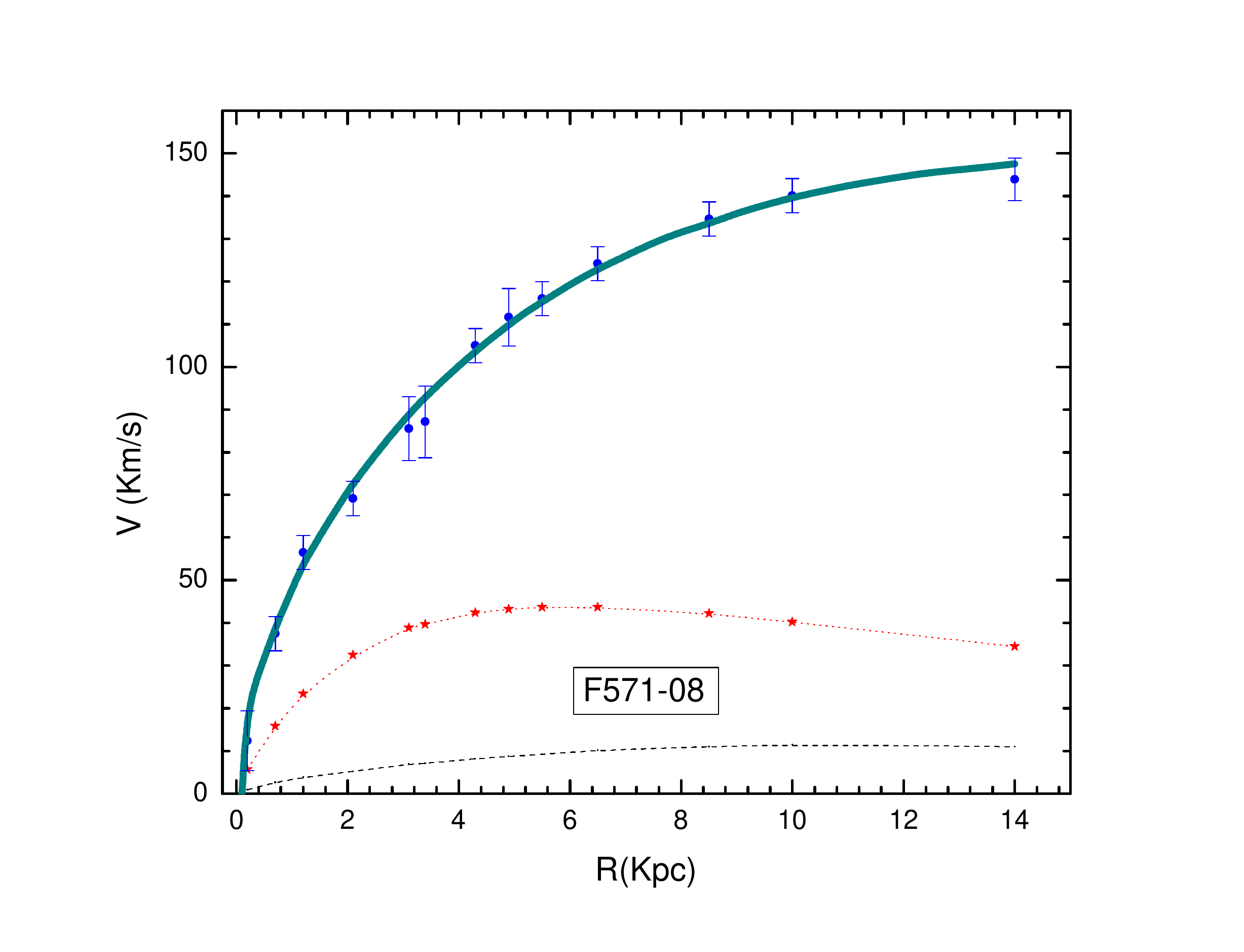}
        } 
        \subfigure[ ]{%

            \includegraphics[width=0.48\textwidth]{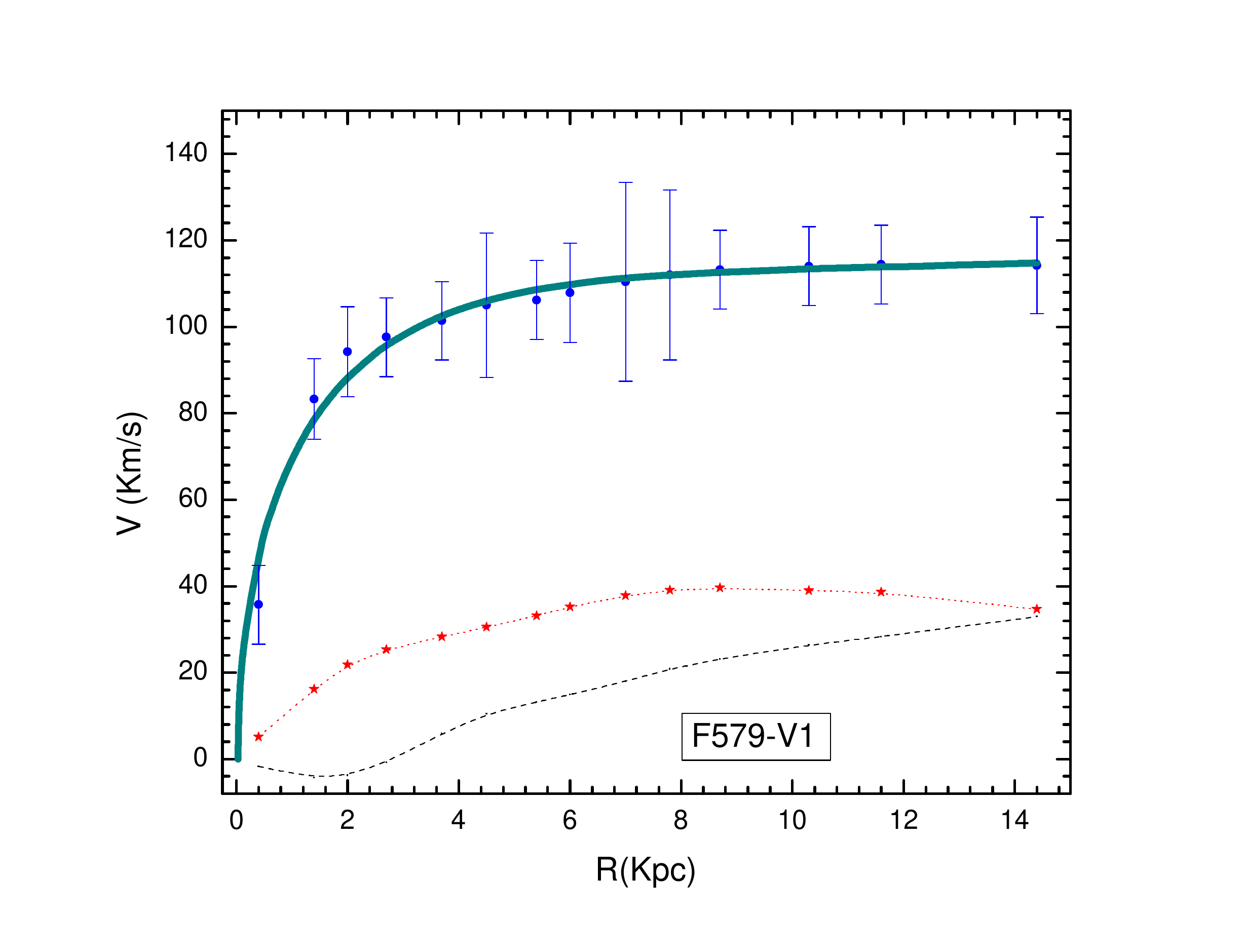}
        }%
        \subfigure[ ]{%

            \includegraphics[width=0.48\textwidth]{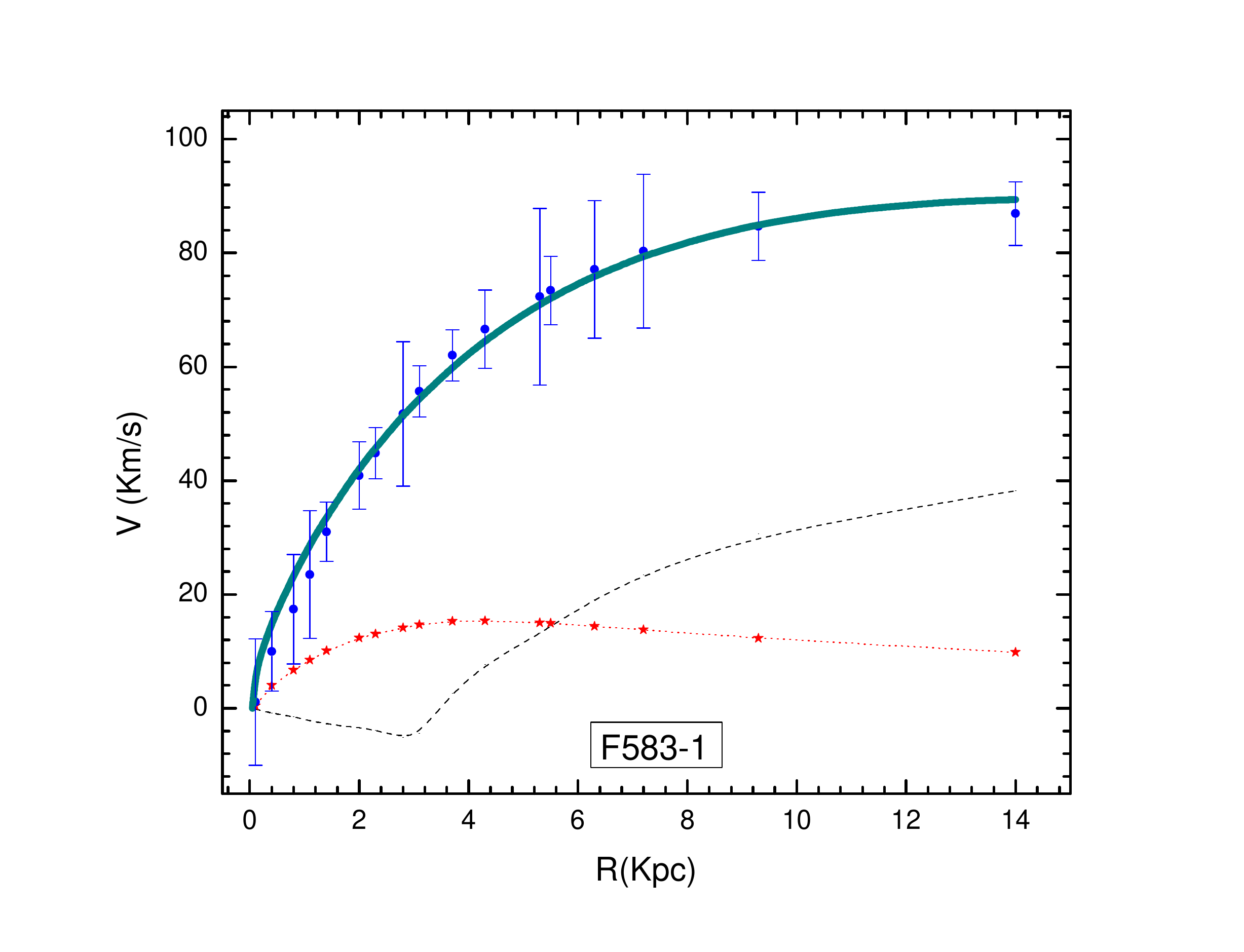}
        }%

    \end{center}
    \caption{%
        The rotation curves of smoothed hybrid HI and H-alpha measurements. The points with the solid lines represent the model prediction. The velocities of gas extended far the galactic disk are the dashed lines. The starred lines indicate the  velocity of stars in the disk. Error bars are uncertainties                                                                                                                                                                                                                                                                                                                                                                                                                                 in velocity measurements and take into account measurement, inclination and asymmetry uncertainties \cite{blok}.
     }%
   \label{fig:subfigures}
\end{figure*}

 \begin{figure*}[htp!]
     \begin{center}
        \subfigure[ ]{%
            \label{fig:first}
            \includegraphics[width=0.48\textwidth]{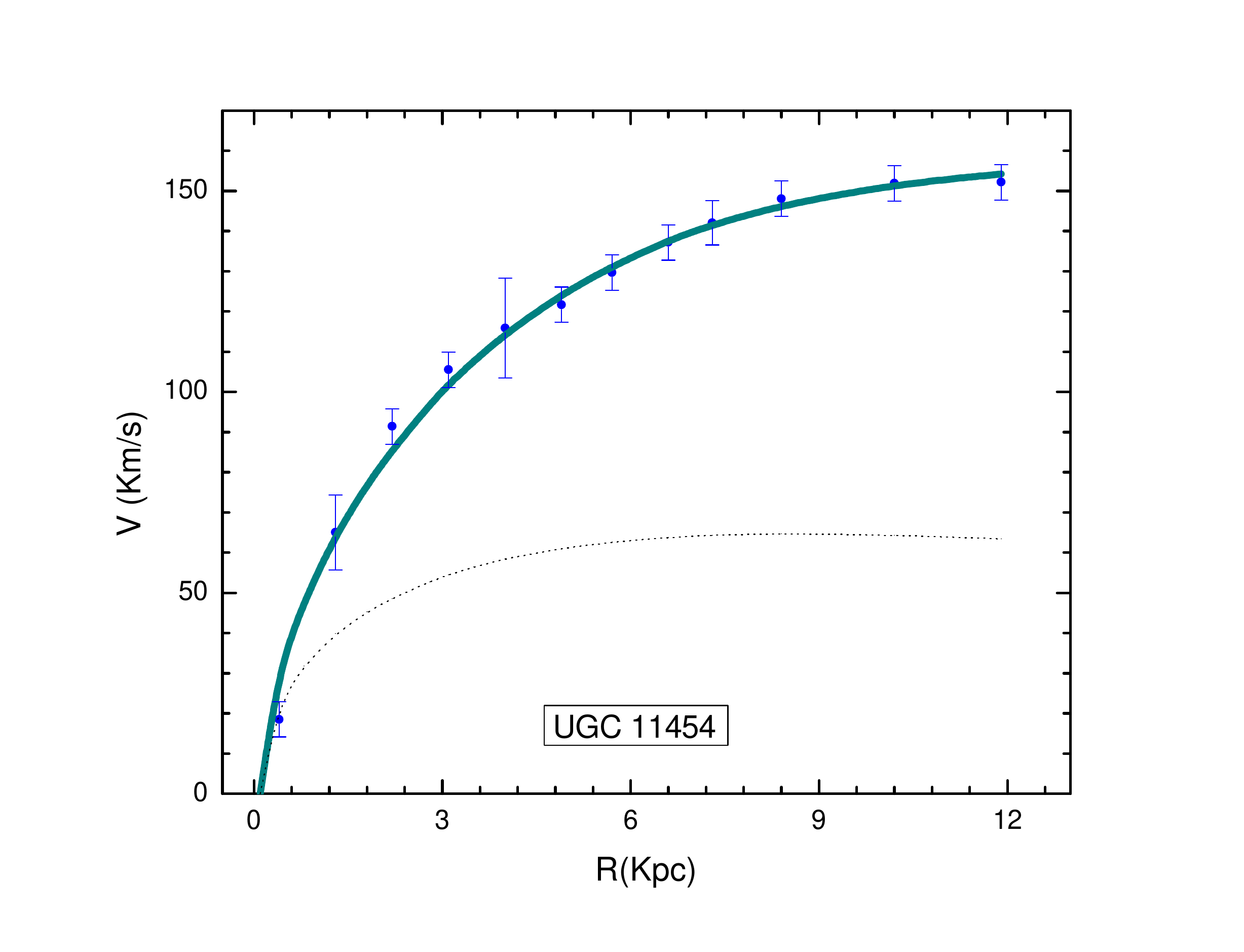}
        }%
        \subfigure[ ]{%
           \label{fig:second}
           \includegraphics[width=0.46\textwidth]{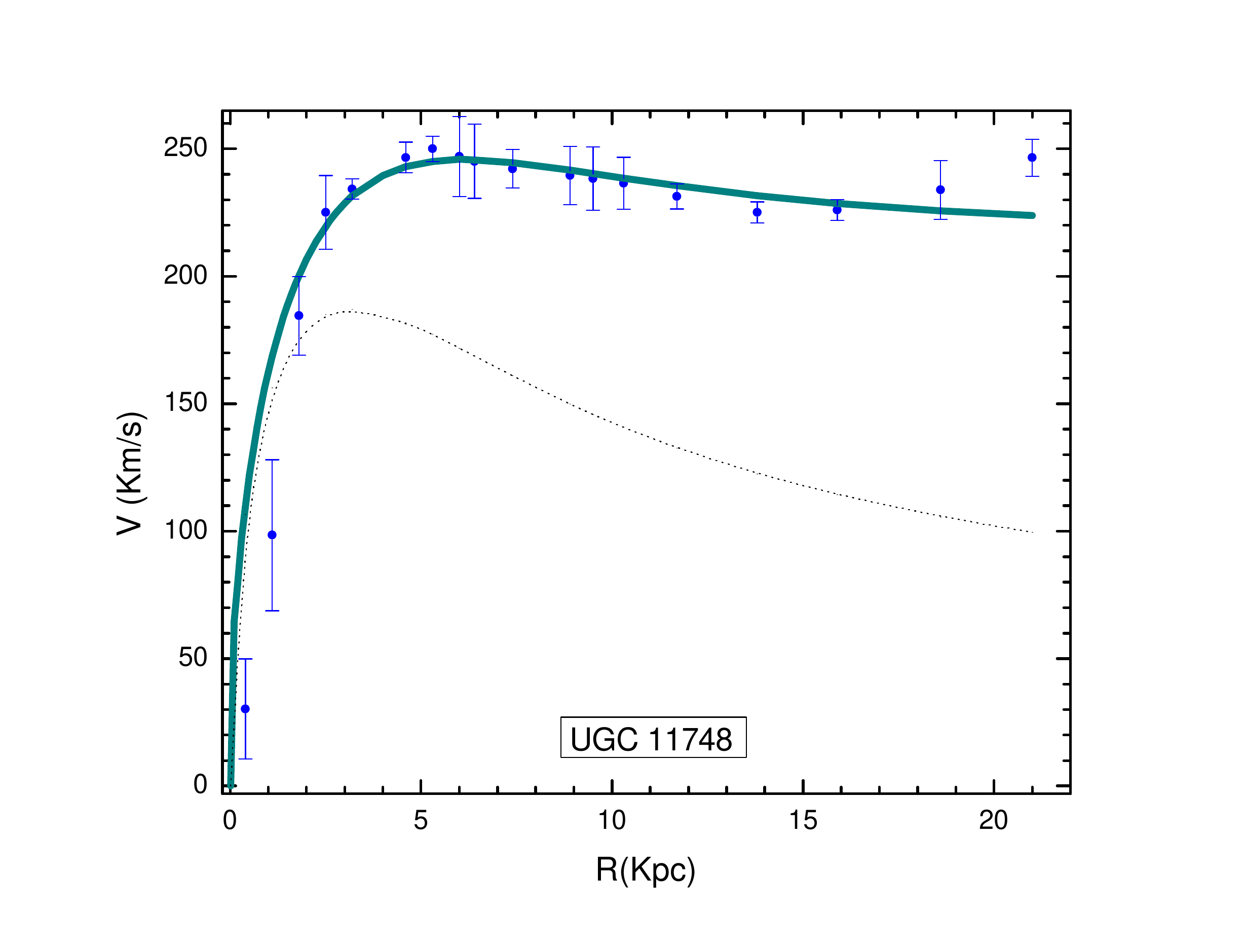}
        }\\ 
        \subfigure[ ]{%
            \label{fig:third}
            \includegraphics[width=0.46\textwidth]{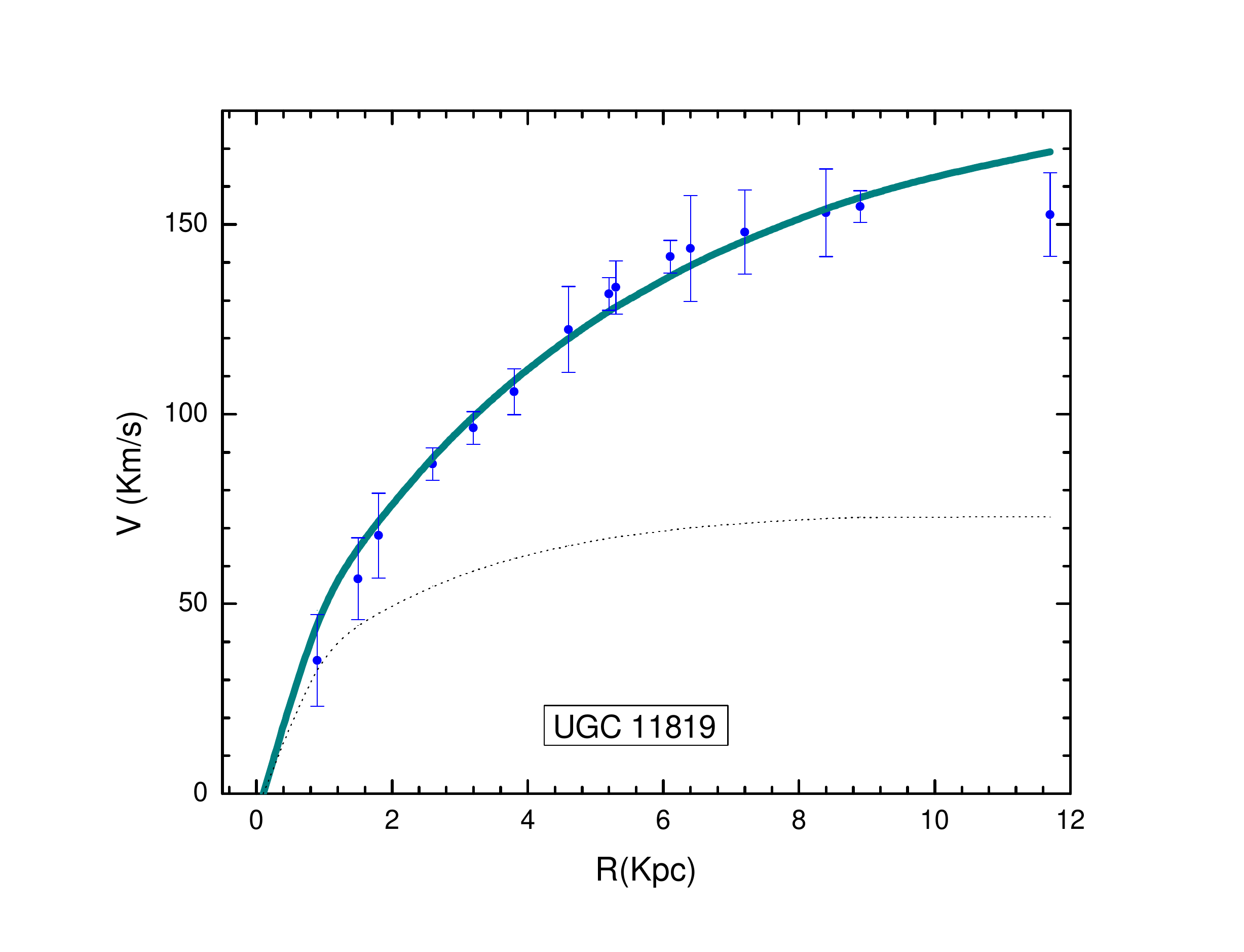}
        }%
        \subfigure[ ]{%
            \label{fig:fourth}
            \includegraphics[width=0.46\textwidth]{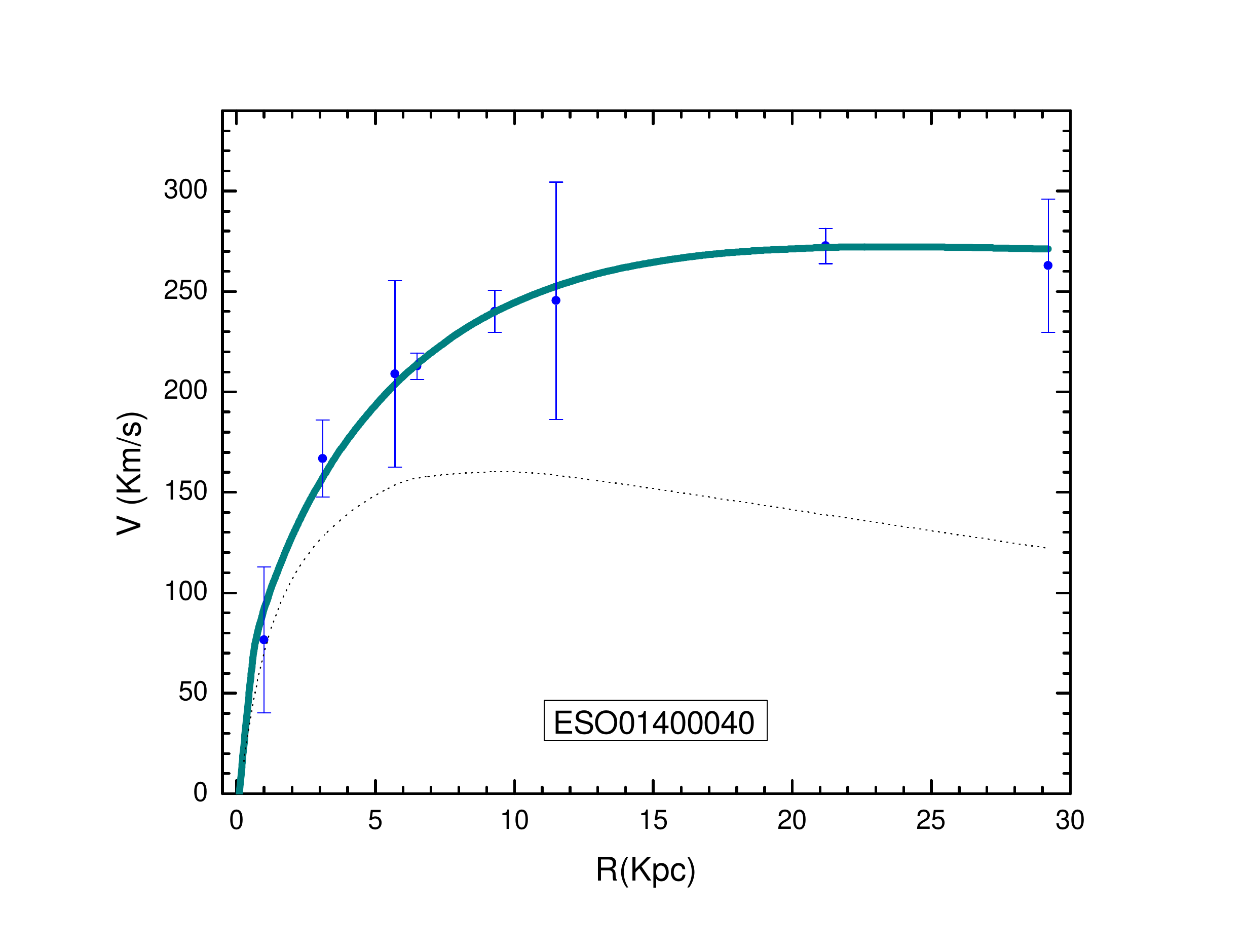}
        }%

\subfigure[ ]{%
            \label{fig:first}
            \includegraphics[width=0.46\textwidth]{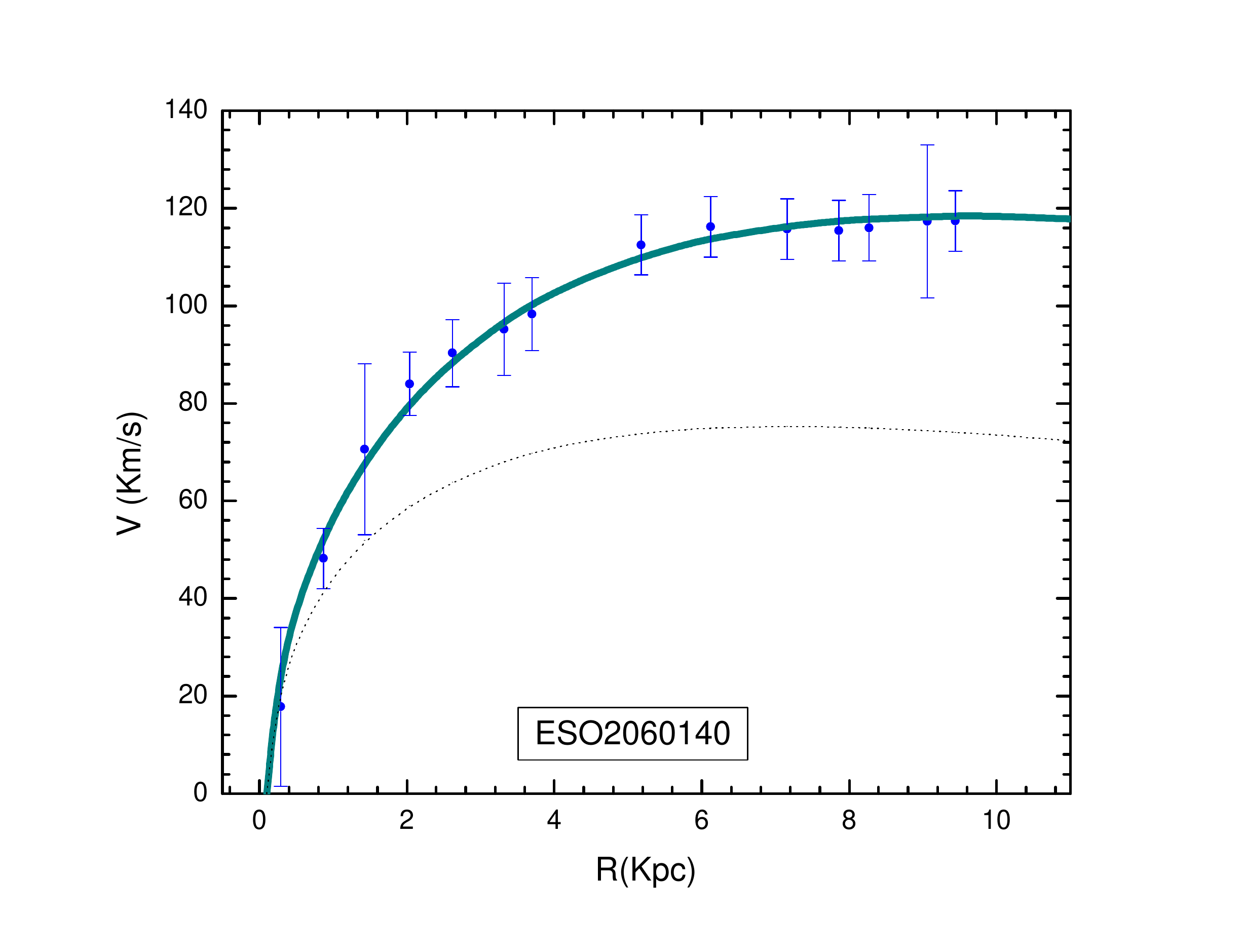}
        }%
\subfigure[ ]{%
            \label{fig:first}
            \includegraphics[width=0.46\textwidth]{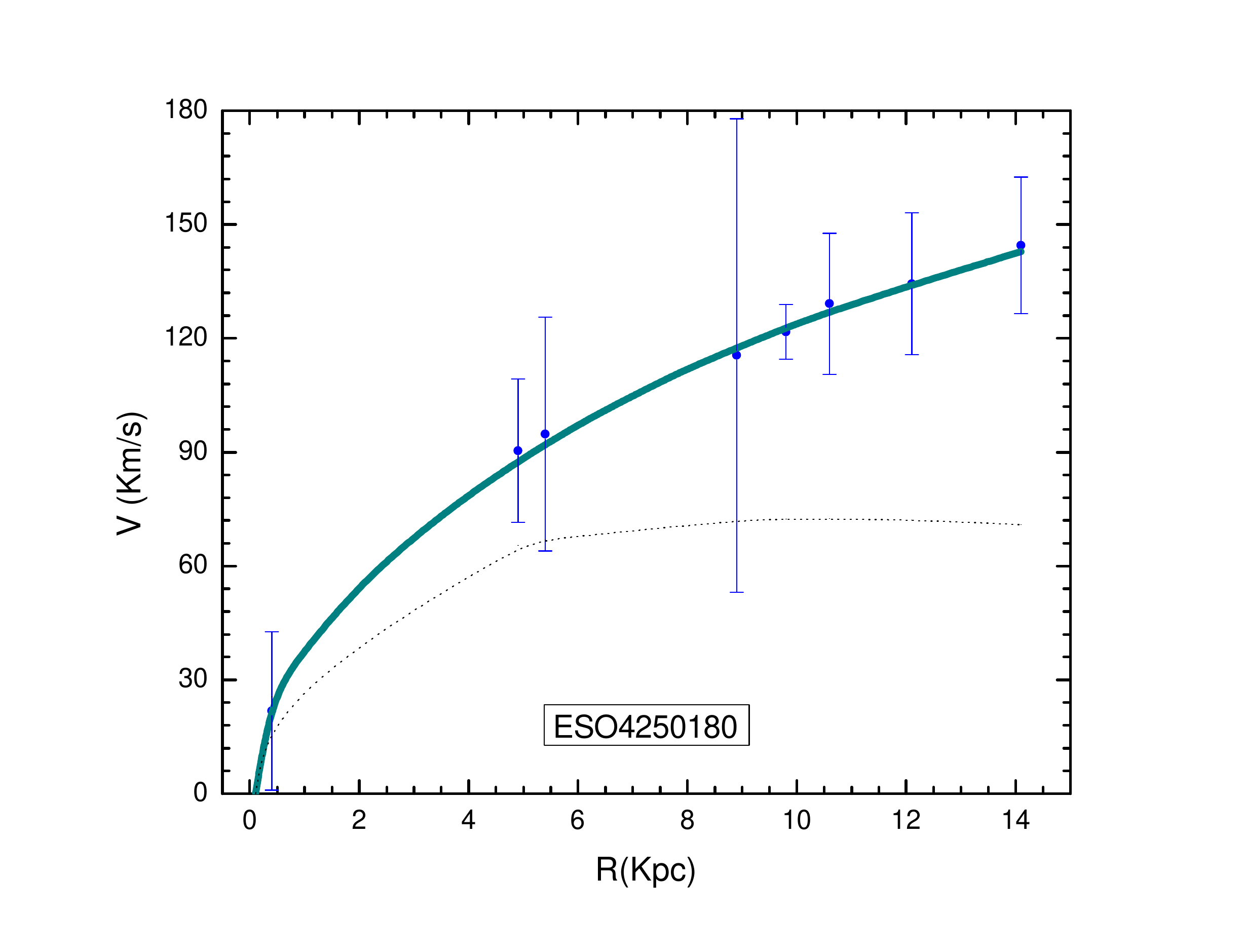}
        }%

    \end{center}
      \caption{%
        Rotation curves of the second group of 6-LSB galaxies. The solid lines represent the model predictions and the error bars represent the observed values. The dashed lines are the rotation curve without the extrinsic contribution.
     }%
\end{figure*}

 \begin{figure*}[htp!]
     \begin{center}
        \subfigure[ ]{%
            \label{fig:first}
            \includegraphics[width=0.45\textwidth]{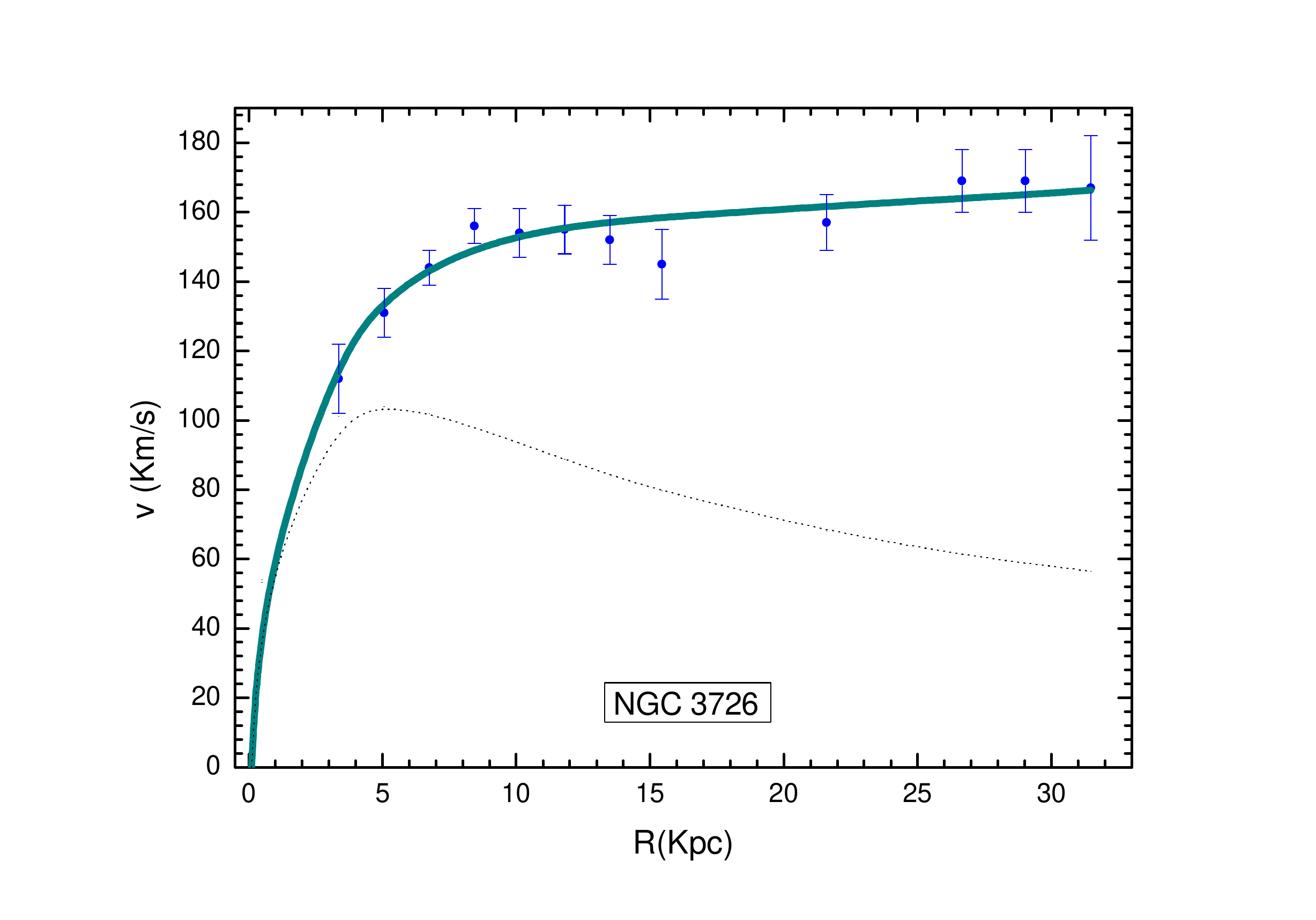}
        }%
        \subfigure[ ]{%
           \label{fig:second}
           \includegraphics[width=0.43\textwidth]{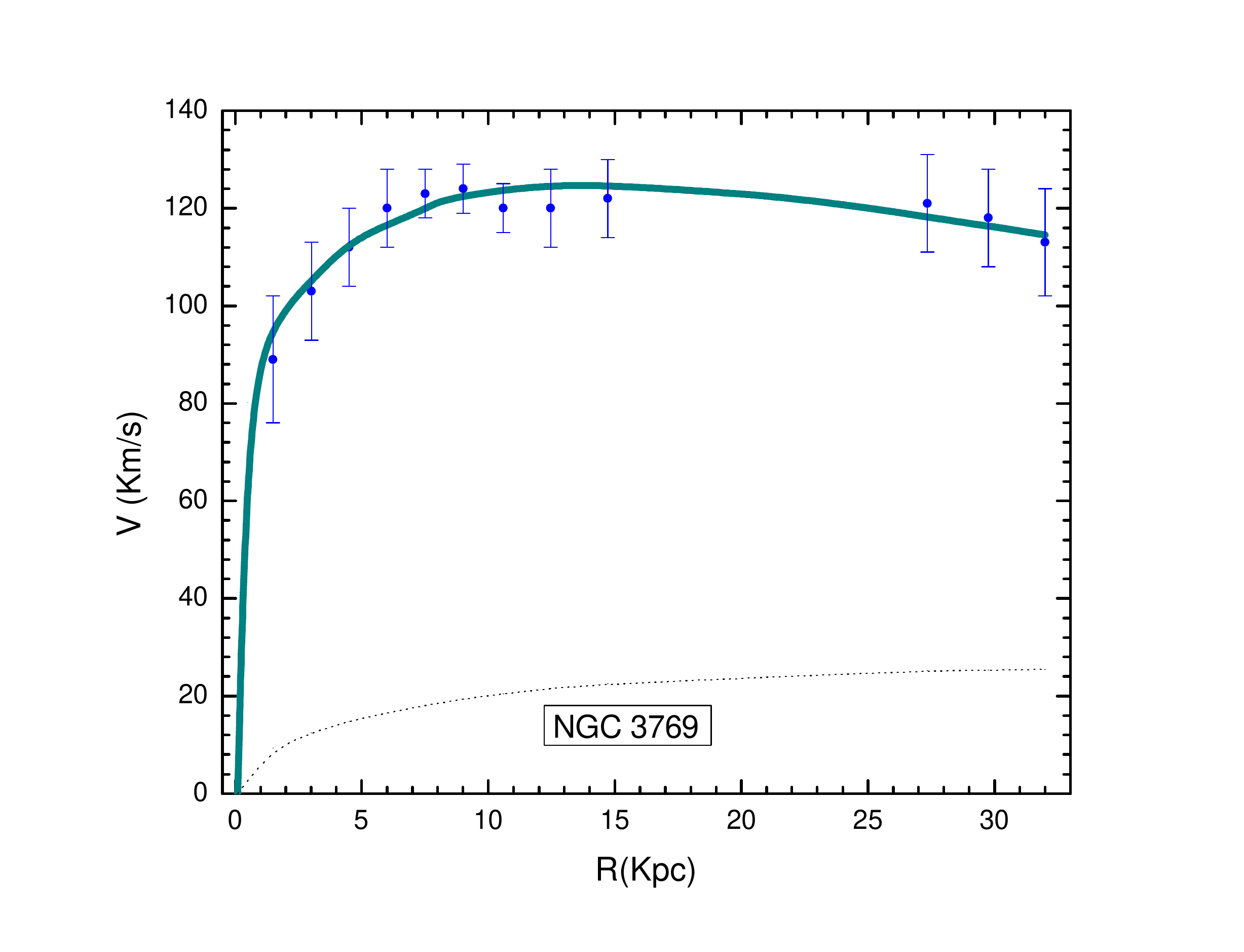}
        }\\ 
        \subfigure[ ]{%
            \label{fig:third}
            \includegraphics[width=0.45\textwidth]{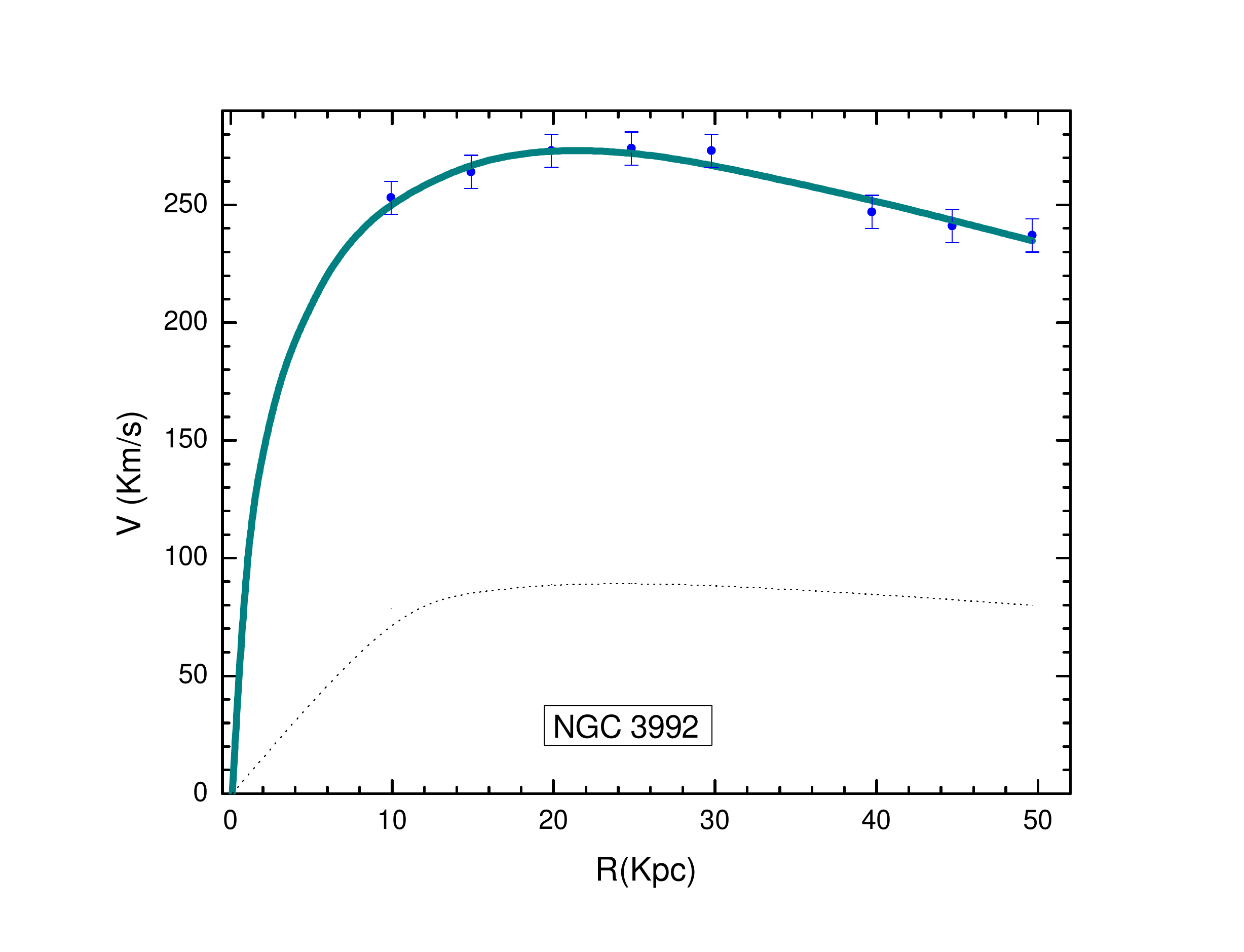}
        }%
        \subfigure[ ]{%
            \label{fig:fourth}
            \includegraphics[width=0.45\textwidth]{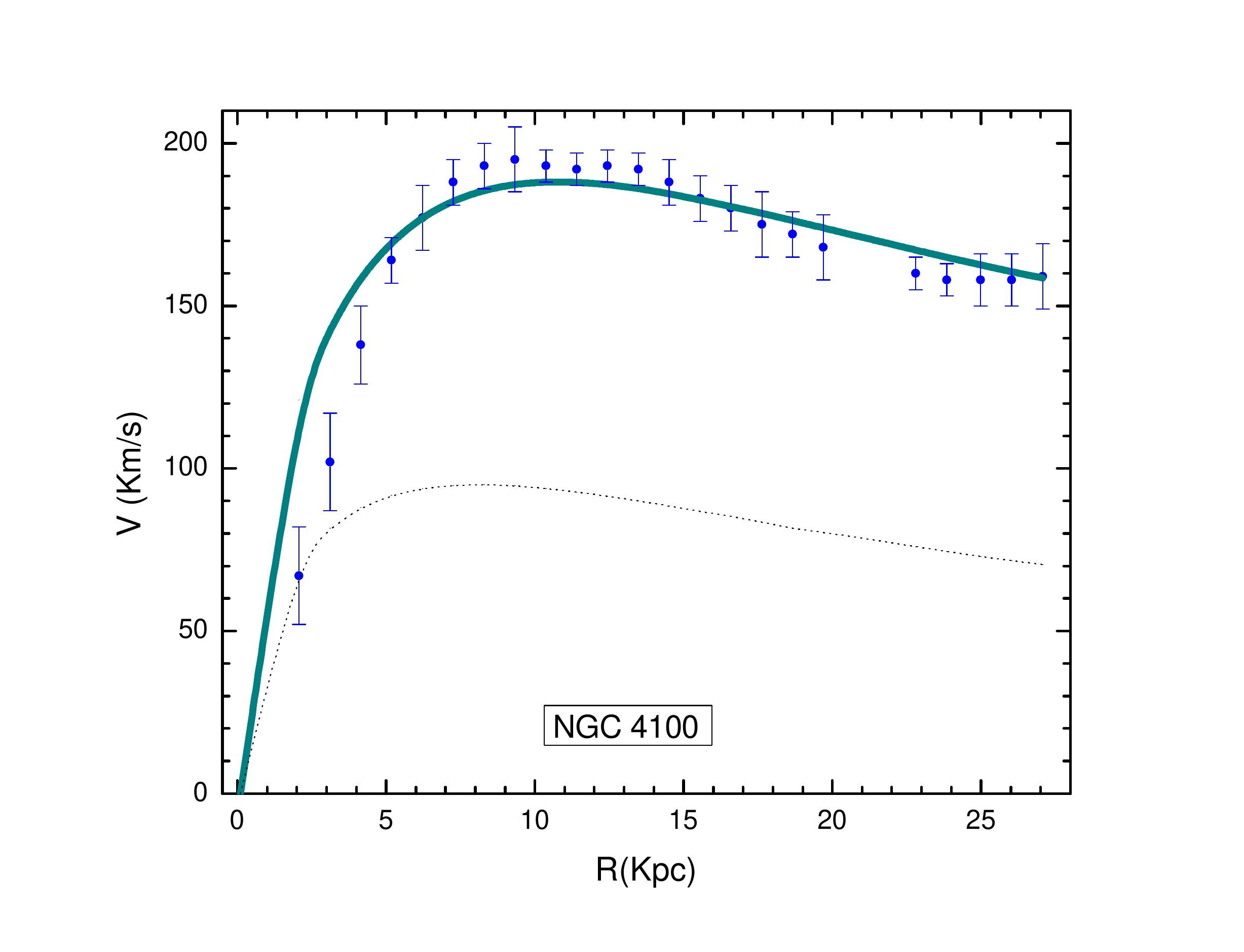}
        }%

\subfigure[ ]{%
            \label{fig:first}
            \includegraphics[width=0.45\textwidth]{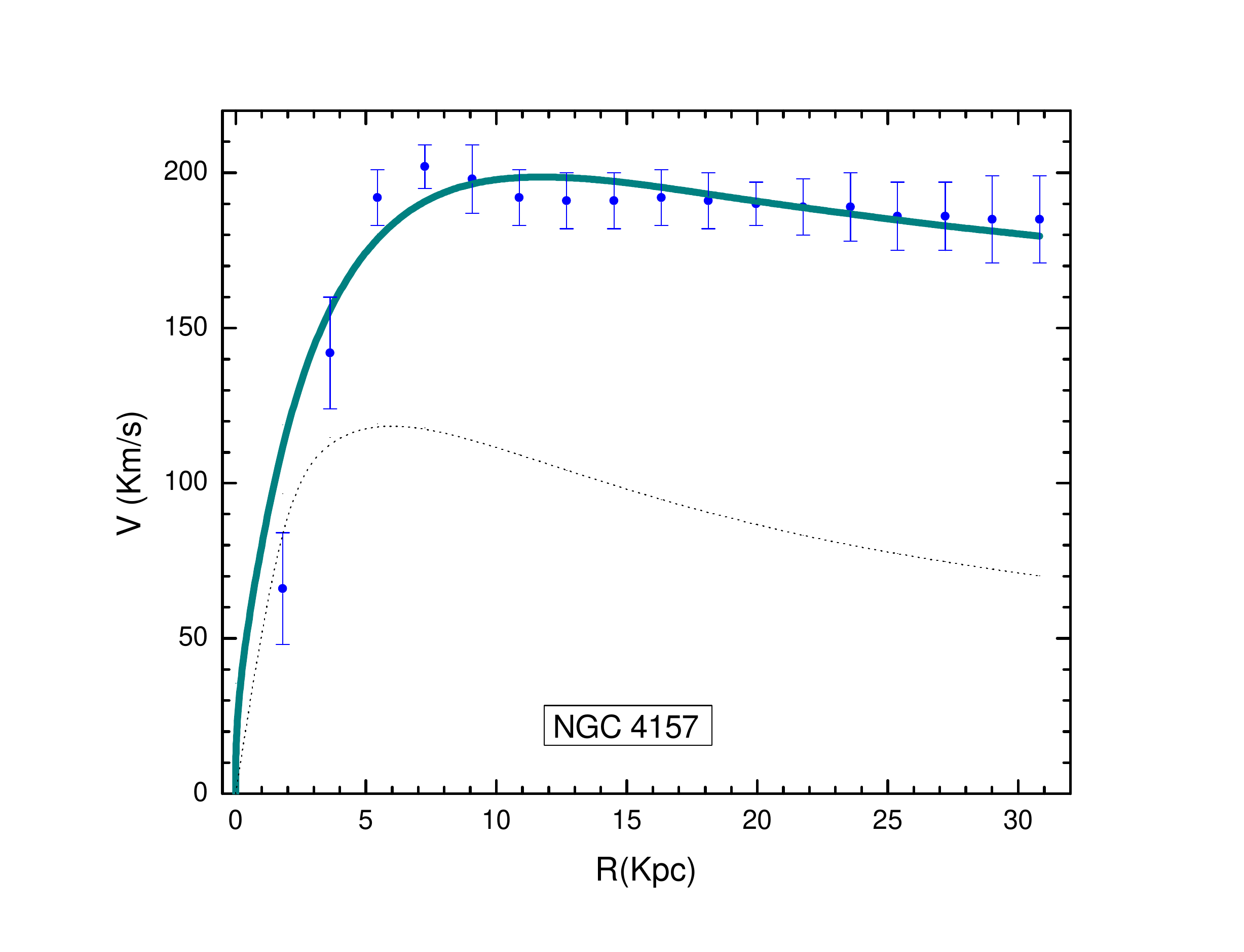}
        }%
\subfigure[ ]{%
            \label{fig:first}
            \includegraphics[width=0.45\textwidth]{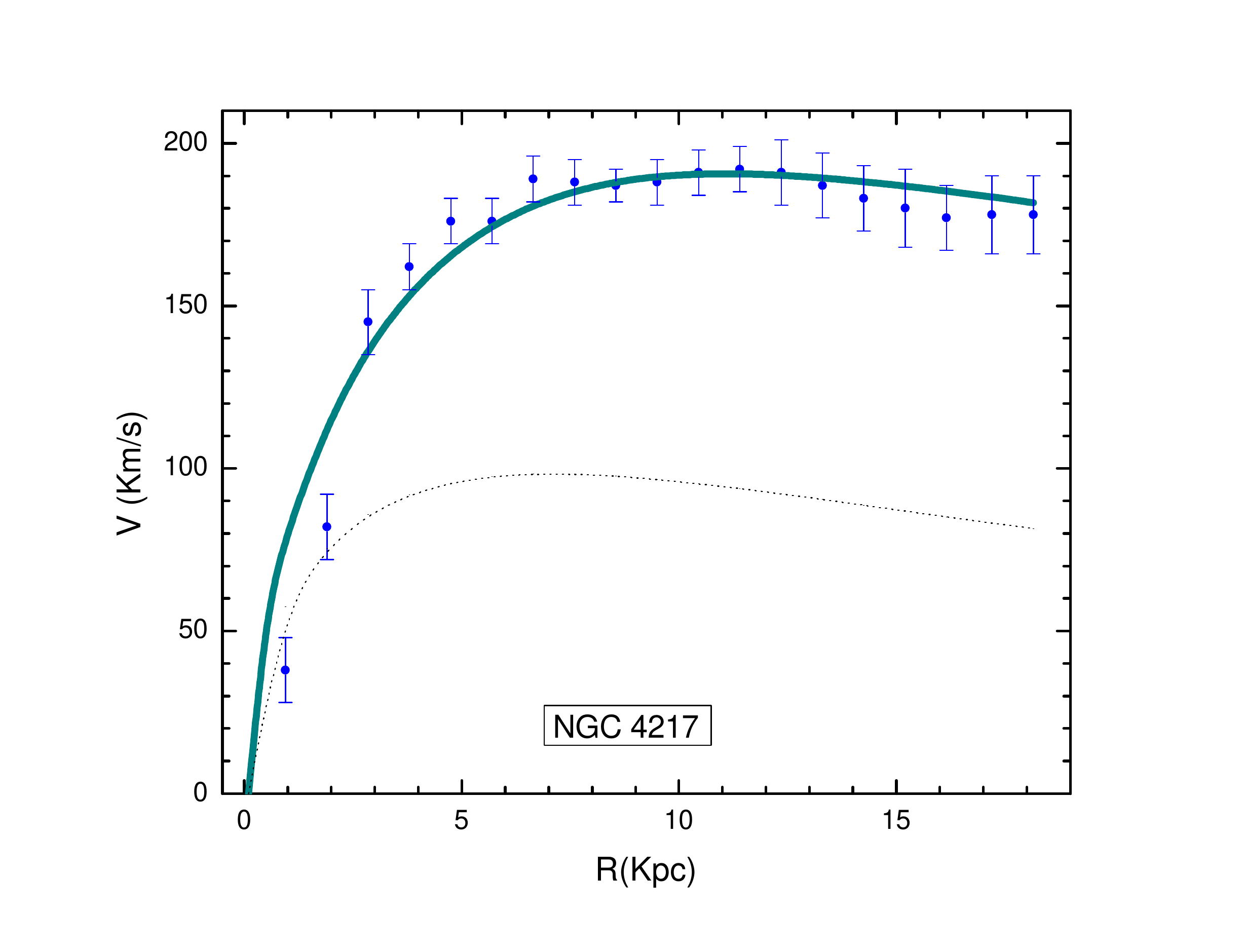}
        }%

    \end{center}
       \caption{%
        Rotation curves of 6-HSB galaxies. The solid lines represent the model predictions and the error bars represent the observed values. The dashed lines are the rotation curve without the extrinsic contribution.
     }%
\end{figure*}

The specific analysis by the authors in \cite{verheijen} was based on measurements of the HI 21-cm line synthesis image survey using the Westerbork Synthesis Radio Telescope (WSRT). The mean distance adopted by the authors for the cluster was of 18.6 Mpc. As a result, the parameter values $r_l, \gamma_0, \beta_0$ are presented in table (4) as well as $\chi^2_{red}$ for the 6-HSB galaxies.

Concerning values of the obtained $\chi^2_{red}$, we noted that UGC11454 and NGC4157 are near 1. Thus, these cases would represent only the best-fittings in the sample. Moreover, the critical value for $\chi^2_{red}$ would represent a failure of the model mainly on the ESO4250180 galaxy. Accordingly, the model could not be used for the rotation curves problems even with the fits of rotation curves presented in figures (1) and (2) are very closed to the observed ones.

On the other hand, we must analyze the obtained values of $\chi^2_{red}$ with caution. Small values of $\chi^2_{red}$ does not indicate a poor model, but the uncertainties were conservatively overestimated. The authors in \cite{blok} reported that error bars can easily dominate any model fit and constrain the $\chi^2_{red}$ values, consequently the goodness-of-fit parameters must assume values smaller than 1 generally. This is due to the difficulty to obtain fair estimates of the errors on many observational quantities. In this sense, $\chi^2_{red}$ values can only tell us relative merits of different models. As a result, this makes the error distribution not representable as a random error in the usual statistical sense. Thus, different smooth procedures may induce unexpected correlations among data point. For instance, eventual misalignment between hybrid HI and H-alpha measurements could induce a bias in data points leading to a non gaussian distribution and maybe the appearance of eventual outliers.

If we analyze our obtained $\chi^2_{red}$ we attend to the constraint proposed in \cite{blok} that states a probability $p$ which measures the data and the model compatibilities.\\

In the next pages, the resulting rotation curves of the sample of 16-galaxies are presented.
\begin{center}
\begin{table*}[htp]
{\small
\hfill{}
\begin{tabular}{*{6}{l}}
\toprule
\bfseries \bfseries Galaxies &  $r_l$ & $\gamma_0$ & $\beta_0$ & $\chi^2_{red}$ & $p$ \\
\midrule
NGC3726                 &   2.768   +/- 0.296    &  0.363   +/- 0.087   &   0.335   +/- 0.124     & 0.634  & 0.7689     \\
NGC3769                 &   19.170  +/- 3.898    &  104.681 +/- 48.950  &   -1.553  +/- 0.072     & 0.185  & 0.9957     \\
NGC3992                 &   13.383  +/- 3.838    &  9.887   +/- 6.361   &   -1.120  +/- 0.235     & 0.390  & 0.8560     \\
NGC4100                 &   4.515   +/- 3.440    &  2.0321  +/- 3.764   &   -0.690  +/- 0.825     & 1.816  & 0.0141     \\
NGC4157                 &   3.193   +/- 0.637    &  0.695   +/- 0.336   &   -0.160  +/- 0.225     & 1.121  & 0.3326    \\
NGC4217                 &   3.944   +/- 8.555    &  1.623   +/- 8.574   &   -0.495  +/- 2.721     & 2.167  & 0.0044     \\
\toprule
\end{tabular}}
\hfill{}
 \caption{Fitting parameters for HSB galaxies with their individuals statistical parameters $\chi^2_{red}$ and the chance $p$  .}
\label{tb:tablename2}
\end{table*}
\end{center}

The values of $p$ in tables (2) and (4) were evaluated using the standard $\chi^2$ test. These values  indicate that if $p> 0.95$ we have a good data plus model fit and $p<0.05$ means an inconclusive analysis (due to systematic effect) or a new model must be found. The ``average'' fit is obtained from a range of $0.05 < p < 0.95$.

As a result, we checked that even if the case of ESO4250\\180 with the lowest obtained $\chi^2_{red}$ the model complies with a good values for $p$. The fitting parameters in most cases present a similar pattern as those shown in \cite{blok}. However, it is worth pointing out that 1-LSB and 2-HSB galaxies presented a peculiar situation. \\

The three galaxies $\;$UGC11748, $\;\;$NGC4100 and NGC4217 have in common a high $\chi^2_{red}$ and $p < 0.05$. Comparing our results to the data \cite{blok}, we obtained the same non-representative value for chance $p$. For UGC11748 we obtained a similar chance $p$ compatible with the data. In both cases, $\chi^{2\;model}_{red} < \chi^{2\;data}_{red}$ indicates that the data were conservatively overestimated.

The NGC4100 galaxy presents strong warps which constrain an accurate measurements. For NGC4217 the presence of a huge bulge, high inclination and HI holes in inner regions \cite{verheijen2} that lead  to a steeper rotation curve. As a result, in this particular group, the systematic effects seems to be dominant which make the goodness-of-fit test inappropriate to define a clearer analysis.

The overall conclusion is that the shape of the rotation curves present in this sample of 16-galaxies was not heavily affected by the overestimated uncertainties and presents a similar pattern as shown in literature \cite{blok,Mannheim2,verheijen,verheijen2}.

\section{Summary and remarks}
In the  course of  our  study,  we  realized that   dark matter   cannot be   explained  only  from the    local observations  of  galaxies  clusters  and  merging clusters. One possible   explanation is that the perturbation is made  through  the   intermediation of the gravitational  field  of  dark matter. If one  asks  for the origins  of  dark matter  we  end  up  in the  early  universe,  when  dark  matter  was   supposed  to  break the  cosmological  homogeneity  of  baryons  thus  creating the  large  structures that  are   observed  today,  such as  galaxies and clusters of galaxies.
In this  context,  the   natural    starting point  to  study  the dark matter gravitational  field  is in  the    epoch of  structure  formation. The local  effects  of the dark matter  gravitational  field  are obtained  from the   local  limit of the  cosmological gravitational  field.

Here  we  have     presented  an  application of   Nash's  theorem on the  formation  of geometrical  structures  by   metric  perturbations as   an  alternative to  the  dark matter gravitational  perturbation.

In a brief  justification,  we  have  discussed that   Nash's theorem  does not only  eliminate  the   ambiguity  of  Riemann's notion
of  curvature,  but it also provides   a  mathematically sound way  to   construct   any    Riemannian  manifold  by continuous
perturbations  along   extra  dimensions. We  have conceded that the  existence of  these  extra  dimensions    does  not  represent a  breakdown  of  the  very   sensible  intrinsic view  of  geometry laid  down  by  Gauss and  Riemann. Rather,  their    existence  improve Riemann's  geometry   in  the sense that  it  restores   the relativity  of  shapes of    manifolds,  replacing  somewhat the absolute notion of  curvature described  by the  Riemann tensor  alone by   a   relative  notion of  curvature.

Given the  Riemann tensor $\;$of$\;$ the $\;$ bulk$\;$ (given $\;$by the Einstein-Hilbert principle), then  the   Riemann tensor of the embedded  space-time acquires   a  reference standard. The  four-dimensionality of  space-time  is regarded here  as an experimental $\;$fact$\;$ related to the  $\;$symmetry $\;$properties$\;$ of  Maxwell's  equations  or, more  generally, of the   gauge theories of the standard  model.  Thus,  any matter that interacts  with  gauge  fields,  including we  as  observers, must remain  confined   to  four-dimensions. However,  the gravitational interaction  does not  fit  in the  same   gauge scheme.   Therefore, and  in accordance  with  Nash's   theorem, gravity  is  regarded   as   different  from gauge  fields,  because it is not  confined,  but  it  is  perturbable   along the extra  dimensions.

The metric  perturbation and  the  embedding lead  naturally  to  a  brane-world-like $\;$higher $\;$ dimensional $\;$  structure  which is
very   similar to  the   brane-world  program,  except that  it is not  string inspired as  suggested by the  adjective ``brane''. Furthermore  it is  completely  independent of the many  existing  brane-world models.

Most of our mathematical  tools   are based  on  strong and  well known  results in  differential geometry. In  particular, the   primary  role of the  extrinsic  curvature in Nash's perturbations represents  an innovation  when  compared  with the traditional  perturbative  methods  in  cosmology. In fact, it  represents  a  rank-2  symmetric   tensor  which   is  independent  of  the metric but induces  the  metric perturbations.

In the  cosmological  side,  we  have  discussed  our  results   with  the present  observational  data. In doing so  we  have   started  with  cosmology,  applying  equations  (\ref{BE1}) and (\ref{BE2}) to   the FLRW  cosmological  model,   finding  that   Friedman's  equation is  modified  by the presence of the extrinsic  curvature.  The  theoretical power  spectrum  closely compares    with  the  latest  experimental result  and  our previous   results  show  that   it also agrees  with  the
accelerated expansion of the universe. In a previous  publication we  have  found that the same   extrinsic  curvature    perturbs Friedman's  equation   in such a way  that it is    consistent  with  the  observed  acceleration of the universe \cite{gde2}. Since  gravitational perturbation means  also geometry perturbation,  the  understanding of   structure  formation  implies in the  study on the generation of  geometrical  forms in geometry.  This  goes  beyond the  classical   classic perturbation  theorems used  in general  relativity, in the  line   described  by  J.  Bardeen \cite{Bardeen},  which  admit  that the Riemann  geometry  of  space-time  is already established  together  with coordinate   gauges  resulting from the  diffeomorphism  invariance  of   the  theory.

All   astrophysical  observations  are  made as  if  we  were  in  the Euclidean  3-space.  However,  when  we  finally   write  a  theory   to describe  those  observations  we use  a  different Riemann geometry. The  difficulty is  that   the   objects  described by the  chosen  geometry (by Einstein), does not necessarily coincide with  the objects  described by  the observation geometry. In other words,  we    cannot be  sure  that  our  theoretically constructed  structures using  Riemann geometry correspond  to  what   is actually  observed.   It appears  that the    recent  astrophysical observations  are  telling  us  about  this  geometrical  difference. We have shown that even we have a flat space or metric, a perturbation can be generated in an embedded space-time. This is essence of Nash's theorem.

Next  we have  applied  the   same    equations (\ref{BE1}) and (\ref{BE2}) to  the  study of  the   local  dark matter gravitational  field  of  galaxies considering a 4-d spherically symmetric metric as a starting point. It is worth noting that a dark matter component was not necessary to invoke. Even with the problematic group of UGC11748, NGC4100 $\;$ and $\;$ NGC4217, where the systematic effect seems to be more strong entangled, the overall result presented in the present sample galaxies were satisfactory with similar rotation curves present in literature. A possible further step is to extend the sample mainly on the HSB galaxies to large one. Giant and dwarf galaxies should be also verified. Eventually, a different geometrical structure can be used to study  possible different effects that such changing can imprint on the results. An oblate spheroid can be used since it naturally carries variations of angles, which can be interesting for dealing with the inclination parameter of galaxies (e.g., NGC4217 presents a high inclination). In order to obtain a more interesting results, we postponed to a later paper a more sophisticated statical analysis as we intend to work with a larger sample.

It is important to point out that the perturbational aspects lie in the rotation velocity obtained which is modified by the extrinsic curvature with the appearance of the term $\left[\gamma_0\left(\frac{r}{r_c}\right)^{\beta_0}\right]$. It is interesting to note that this correction term in eq.(\ref{eq:velocrot2}) resembles a similar expression if one calculates the  rotation velocity for a thin disk using the Weyl coordinates \cite{gautreau} with a potential $\phi\rfloor_{disk}= -\frac{1}{2}(1+ g_{44})\rfloor_{z=0} = -\frac{1}{2}(1-e^{2c_0} r^{\delta_0} )$ where $c_0$ and $\delta_0$ are integration constants. This interpretation seems to be compatible with the fact that the galaxy core is spherically symmetric and the outermost regions are disk-like shape.

There  are  three  related  problems  which  deserve  further  attention  and  are   presently  in progress.  The  first problem   concerns    the  compatibility    with    the  various ad hoc brane-world models,   which  use   dif         ferent  bulk geometries  and  or additional  symmetries.   The  fact  that  eq.(\ref{BE2})  is  a  homogeneous  equation (a consequence of  confinement  condition),  has  forced us to appeal to a conformal flat condition (\ref{eq:flatcondition}).  Our   explanation is  that   the    embedding  equations    have  two    independent   variables   $g_{\mu\nu}$  and  $k_{\mu\nu}$  and   one   missing  equation,  which is  usually  supplied  as  an  additional  assumption in the mentioned  models. Regarding   $k_{\mu\nu}$  as  a    spin-2  field  over the brane-world, then  it  must obey  an  Einstein-like  equation.  We  already  know    what   is  that  equation as stated in a previous communication \cite{gde2} but the  source term  is  still  undefined. The second problem concerns  inflation.  If   the  extrinsic  curvature effect agrees   with the power  spectrum of the PLANCK collaboration, then  we need to  explain  how  it relates  to the  inflaton. In our  view, since our arguments in this paper were all classical, this  can  only be  solved  when a proper  quantum  theory of the brane-world is  developed  along  with  Nash's  theory of   embedded geometries. The  third problem   is  the  explanation of  the cosmological  constant problem  within  the  context of the brane-world.  As  the  reader  may  have  already guessed,   the  extrinsic  curvature    must play  a role  on that specific  problem.

\acknowledgement{We are grateful to Dr. S.S. McGaugh for his prompt answers to questions on the data. We also thank Dr. Luciano Lapas for his valuable help in the programming issues.}


\begin{thebibliography}{00}

\bibitem{Zwicky} F. Zwicky, \emph{Die Rotverschieb ung von extragalaktischen Nebeln}, Helv. Phys. Acta, 6, 110, (1933). [Translated into english and republished in Gen. Rel.Grav., 41, 1, 207-224, (2009).]

\bibitem{rubin}V.C. Rubin, W. Kent Ford Jr., \emph{Rotation of the Andromeda Nebula from a Spectroscopic Survey of Emission Regions}. Astrophys. J., 159, 379, (1970).

\bibitem{Milgrom} M. Milgrom, \emph{A modification of the Newtonian dynamics as a possible alternative to the hidden mass hypothesis}, Astrophys. J. 270, 365, Ibid p.371, Ibid, p.384, (1983).

\bibitem{Moffat} J.R. Bownstein, J.W. Moffat, \emph{Galaxy rotation curves without nonbaryonic dark matter}, Astrophys. J., 636, 721, (2006).

\bibitem{Bekenstein} J. Bekenstein, M. Milgrom, \emph{Does the missing mass problem signal the breakdown of Newtonian gravity?}, Astrophys. J., 286 , 7, (1984).

\bibitem{Freese}  K. Freese,  \emph{Cardassian Expansion: dark energy density from modified Friedmann equations}, New Astron.Rev., 49, 103, (2005).

\bibitem{Skordis} C.  Skordis et. al., \emph{Large Scale Structure in Bekenstein's Theory of Relativistic Modified Newtonian Dynamics}, Phys. rev. Lett. 96, 011301, (2006).

\bibitem{Dodelson}  S.  Dodelson, M. Liguori, \emph{Can Cosmic Structure Form without Dark Matter?}, Phys. Rev. Lett. 97,  231301, (2006).


\bibitem{Mannheim}  P. D. Mannheim,  \emph{Alternatives to dark matter and dark energy}, Prog.Part.Nucl.Phys., 56, 340, (2006).

\bibitem{Mukhanov}  V.  F. Mukhanov, H. A. Feldman, R. H.  Brandenberger, \emph{Theory of cosmological perturbations}, Phys. Rep. 5,6,  203-333, (1992).

\bibitem{Padmanabham}  T.  Padmanabham,   \emph{Structure  Formation in the Universe}, Cambridge  U. press, (1993).

\bibitem{Hooper} D. Hopper, E. A. Blatz,  \emph{Strategies for Determining the Nature of Dark Matter}. ArXiv:0802.0702V2.

\bibitem{Marco}M. Taoso et al, \emph{Dark Matter Candidates: A Ten-Point Test}. ArXiv:0711.4996v2.

\bibitem{capo}S. Capozziello, G. Lambiase, \emph{A comprehensive view of cosmological Dark Side}, ArXiv:1304.5640v1.

\bibitem{Nash}J. Nash, \emph{The imbedding problem for Riemannian manifolds}, Ann. Maths., 63, 20, (1956).

\bibitem{GDE}  M.D. Maia et al, \emph{On the geometry of dark energy}, Class.Quantum Grav., 22, 1623,(2005).

\bibitem{Bardeen}  J. M. Bardeen, \emph{Gauge-invariant Cosmological Perturbations}, Phys. Rev., D22:1882, 1905, (1980).

\bibitem{Geroch}  R.  Geroch, \emph{Limits of spacetimes}, Comunn. Math. Phys., 13, 180,(1969).

\bibitem{Walker} J. M. Stewart, M. Walker, \emph{Perturbations of Space-Times in General Relativity}, Proc. Roy. Soc. (London), A341, 49, (1974).

\bibitem{Greene} R. Greene, \emph{Isometric embeddings of Riemannian and pseudo-Riemannian manifolds}, Memoirs Amer. Math. Soc., 97, (1970).

\bibitem{Riemann} B. Riemann,\emph{On the Hypotheses that Lie at the Bases of Geometry (1854)}, (Translated by  W. K. Clifford), Nature, 8, 114 and 136, (1873).

\bibitem{Schlaefli}L. Schlaefli,(Nota alla memoria del. Sig. Beltrami), \emph{Sugli spazzi di curvatura constante}, Ann. di mat., 2 serie, 5, 170, (1871-1873).

\bibitem{Cartan} E. Cartan, \emph{Sur la possibilité de plonger un espace rimannian donné dans un espace euclidien}, Ann. Soc. Pol. Mat., 6, 1, (1927).

\bibitem{Janet} M. Janet, \emph{Sur la possibilité de plonger un espace Riemannien donné dans un espace Euclidien},  Ann. Soc. Polon. Math., 5, 38, (1926).

\bibitem{Campbell} J. E. Campbell, \emph{A Course of Differential Geometry}, Claredon Press, Oxford (1926).

\bibitem{Dahia} F. Dahia, C. Romero, \emph{The Embedding of Spacetime into Cauchy Developments}, Braz.J.Phys., 35, 1140, (2005).

\bibitem{QBW}  M.D. Maia et al, \emph{Brane-world quantum gravity}, JHEP, 04, 047, (2007).


\bibitem{ruba} V.A. Rubakov, M.E. Shaposhnikov, \emph{Do we live inside a domain wall?}. Phys. Lett. B 125, 136, (1983).


\bibitem{ADD} N. Arkani-Hamed et al, \emph{The Hierarchy problem and new dimensions at a millimeter}, Phys. Lett., B429, 263, (1998).

\bibitem{RS}L. Randall, R. Sundrum, \emph{Large Mass Hierarchy from a Small Extra Dimension}, Phys. Rev. Lett., 83, 3370,(1999).

\bibitem{RS1}L. Randall, R. Sundrum,  \emph{ An Alternative to Compactification}, Phys. Rev. Lett., 83, 4690, (1999).

\bibitem{Israel}W. Israel, \emph{Singular hypersurfaces and thin shells in general relativity}, Il Nuovo Cimento, 44, 2, 1-14, (1966).

\bibitem{Brandon}R. A. Battye, B. Carter, \emph{Generic junction conditions in brane-world scenarios}, Phys. Lett. B, 509, 331 (2001).

\bibitem{maia2}M.D. Maia, E.M. Monte, \emph{Geometry of brane-worlds}. Phys. Lett. A, 297, 2, 9-19, (2002).

\bibitem{sepangi} M. Heydari-Fard, H. R. Sepangi, \emph{Anisotropic brane gravity with a confining potential}, Phys.Lett., B649, 1-11, (2007).

\bibitem{sepangi1}S. Jalalzadeh et al, \emph{Classical tests in brane gravity}, Class.Quant.Grav., 26,155007, (2009).

\bibitem{maeda} K. Maeda et al, \emph{The Einstein equations on the 3-brane world}. Phys. Rev. D62, 024012, (2000).

\bibitem{maartens}R. Maartens, \emph{Brane-world gravity}, gr-qc/0312059.

\bibitem{anderson} E. Anderson, R. Tavakol. \emph{Reformulation and Interpretation of SMS Braneworld}, Class.Quant.Grav., 20, L267, (2003).

\bibitem{sahni} V. Sahni, Y. Shtanov, \emph{Braneworld models of dark energy}, astro-ph/0202346.

\bibitem{Tsujikawa} S. S. Tsujikawa et al, \emph{Observational constraints on braneworld inflation: the effect of a Gauss-Bonnet term},  Phys. Rev., D70,063525, (2004).

\bibitem{gong} Y. Gong, Chang-kui Duan, \emph{Supernova constraints on alternative models to dark energy}, astro-ph/0401530.


\bibitem{eisenhart}L.P. Eisenhart, \emph{Riemannian Geometry}, Princeton U.P., 8th ed., New Jersey, (1997).


\bibitem{Crampin}M. Crampin, F.A.E. Pirani, \emph{Applicable  Differential Geometry},  Cambridge  U.P.,  New York, (1986).

\bibitem{capis}A. J. S. Capistrano, M.D. Maia. \emph{Applications of Nash's theorem to cosmology}. In: Antonio Alfonso-Faus. (Org.). Aspects of Today's Cosmology. 01ed. Rijeka: InTech, 2011,133-152,(2011).

\bibitem{Rosen} N. Rosen, \emph{A Particle at Rest in a Static Gravitational Field}, Rev. Mod. Phys., 21, 503, (1949).

\bibitem{binetruy}P. Binetruy et. al., \emph{Non-conventional cosmology from a brane-universe}, Nucl. Phys. B565, 269, (2000); P. Binetruy et al, \emph{Brane cosmological evolution in a bulk with cosmological constant}, Phys. Lett. B477, 285 (2000); J. M. Cline et al, \emph{Cosmological Expansion in the Presence of an Extra Dimension}, Phys. Rev. Lett. 83, 4245 (1999); S. Tsujikawa, A. R. Liddle, \emph{Constraints on braneworld inflation from CMB anisotropies}, JCAP, 0403, 001, (2004); C. Ringeval et.al., \emph{CMB anisotropies from vector perturbations in the bulk}, hep-th/0307100.


\bibitem{gde2}M.D. Maia et.al., \emph{The deformable universes}, Gen. Rel. Grav., 10, 2685, (2011).

\bibitem{planck}P.A.R. Ade et.al,(Planck collaboration), \emph{Planck 2013 results. XVI: Cosmological parameters}, arXiv:1303.5076v1.

\bibitem{Puzia}  T.H. Puzzia et al, \emph{Evidence for the Evolution of Young Early-Type Galaxies in the GOODS/CDF-S Field}, ArXiv:0705.4092.

\bibitem{deVries} N. de Vries et al, \emph{Massive galaxies with very young AGN}, ArXiv:0708.2672.

\bibitem{kasner}E. Kasner, \emph{The impossibility of Einstein fields immersed in flat space of five dimensions}, Am. Journ.Math., 43, 126, (1921).

\bibitem{kasner2}E. Kasner, \emph{Finite representation of the solar gravitational field in flat space of six dimensions}, Am. Journ.Math., 43, 130, (1921).


\bibitem{fronsdal}C. Fronsdal, \emph{Completion and Embedding of the Schwarzschild Solution}, Phys.Rev., 116, 778, (1959).

\bibitem{kruskal}M.D. Kruskal, \emph{Maximal Extension of Schwarzschild Metric}, Phys.Rev., 119, 1743 (1960).

\bibitem{Szekeres}G. Szekeres, \emph{On the singularities of a Riemannian manifold}, Publ. Math. Debrecen, 7, 285, (1960).


\bibitem{blum}G.R. Blumenthal et al, \emph{Contraction of dark matter halos due to a baryonic infall}. Ap.J, 301, 27, (1984).

\bibitem{gautreau} R. Gautreau, R. B. Hoffman, \emph{Exact solutions of the Einstein vacuum field equations in Weyl co-ordinates}, ll Nuovo Cimento B, 61, 2, 411, (1969).

\bibitem{blok} W.J.G. de Blok, S.S. McGaugh, V.C. Rubin, \emph{High-resolution rotation curves of low surface brightness galaxies: Mass Models}, AJ, 122, 2396, (2001).


\bibitem{Mannheim2}P.D. Mannheim, J.G. O'Brien, \emph{Fitting galactic rotation curves with conformal gravity and a global quadratic potential}, Phys. Rev., D85, 124020, (2012).


\bibitem{verheijen}M.A.W. Verheijen, R. Sancisi, \emph{The Ursa Major cluster of galaxies IV. HI synthesis observations}, A$\&$A, 370, 765, 867, (2001).

\bibitem{verheijen2}R.H. Sanders, M.A.W. Verheijen, \emph{Rotation curves of Ursa Major galaxies in the context of Modified Newtonian Dynamics}, AJ, 503, 97, (1998).


\end{thebibliography}
\end{document}